\documentclass[pra,showpacs,twocolumn]{revtex4-1}

\usepackage{amsmath}
\usepackage{amssymb}
\usepackage{hyperref}

\hypersetup{
  colorlinks=true,linkcolor=blue,citecolor=blue,
  filecolor=blue,urlcolor=blue,breaklinks=true
}

\RequirePackage{color}

\begin{document}

\title{Lorentz-violating effects in the spin-1/2 Aharonov-Casher problem}

\author{Edilberto O. Silva}
\email{edilbertoo@gmail.com}
\affiliation{
  Departamento de F\'{i}sica,
  Universidade Federal do Maranh\~{a}o, Campus
  Universit\'{a}rio do Bacanga,
  65085-580 S\~{a}o Lu\'{i}s-MA, Brazil
}

\author{Fabiano M. Andrade}
\email{fmandrade@uepg.br}
\email{f.andrade@ucl.ac.uk}
\affiliation{
  Departamento de Matem\'{a}tica e Estat\'{i}stica,
  Universidade Estadual de Ponta Grossa,
  84030-900 Ponta Grossa-PR, Brazil
}
\affiliation{
  Department of Computer Science,
  University College London,
  London WC1E 6BT, United Kingdom
}

\date{\today}

\begin{abstract}
The effects of a Lorentz symmetry violating background vector on
the Aharonov-Casher bound and scattering scenarios is
considered.
Using an approach based on the self-adjoint extension method, an
expression for the bound state energies is obtained in terms of
the physics of the problem by determining the self-adjoint
extension parameter.
We found that there is an additional scattering for any value of
the self-adjoint extension parameter and bound states for
negative values of this parameter.
By comparing the bound state and scattering results the
self-adjoint extension parameter is determined.
Expressions for the bound state energies, phase-shift and the
scattering matrix are explicitly determined in terms of the
self-adjoint extension parameter.
The expression obtained for the scattering amplitude reveals
that the helicity is not conserved in this scenario.
\end{abstract}


\maketitle

\section{Introduction}
\label{sec:introduction}

The standard model extension (SME)
\cite{PRD.1997.55.6760,PRD.1998.58.116002,PRD.2004.69.105009} has been
an usual framework for investigating signals of Lorentz violation in
physical systems and has inspired a great deal of investigations in this
theme in recent years.
The interest in this issue appears in the different contexts, such as
field theory
\cite{PRD.2014.90.065003,PRD.2014.89.105019,EPJC.2014.74.2799,
PRD.2013.87.047703,PRL.1999.82.3572,PRL.1999.83.2518,
PRD.2001.63.105015,PRD.2006.73.65015,PRD.2009.79.123503,
PD.2010.239.942,PRD.2012.86.065011,PRD.2008.78.125013,PRD.2011.84.076006,
EPL.2011.96.61001,PRD.2012.85.085023,PRD.2012.85.105001,
PRD.2011.84.045008}, gravitation
\cite{PRD.2014.90.025007,PRD.2013.88.025005}, aspects on the gauge
sector of the SME
\cite{NPB.2003.657.214,PRD.1998.59.25002,PRD.2003.67.125011,
PRD.2009.80.125040},
quantum electrodynamics
\cite{EPJC.2014.74.2875,IJMPA.2014.29.1450112,PRD.2012.86.045003,
PRD.2010.81.105015,EPJC.2012.72.2070},
nonrelativistic quantum dynamics and topological phase
\cite{PRD.2014.90.025026,AoP.2013.333.272,EPL.2013.101.51005,
ADP.2011.523.910,JPG.2012.39.55004,JMP.2011.52.063505,PRD.2011.83.125025},
and astrophysics
\cite{PRD.2002.66.081302,AR.2009.59.245,PRD.2011.83.127702}.
These many contributions have elucidated effects induced by Lorentz
violation and the SME has also been used as a framework to propose
Lorentz violating in CPT probing experiments, which have amount to the
imposition of stringent upper bounds on the Lorentz-violating (LV)
coefficients \cite{RMP.2011.83.11,PRL.2002.88.090801,PRL.2000.84.1381}.

The physical properties of the physical systems can be accessed by
including in all sectors of the minimal standard model LV terms.
The LV terms are generated as vacuum expectation values of tensors
defined in a high energy scale.
By carefully analyzing the sectors of the SME some authors have
specialized in introducing news nonminimal couplings between fermionic
and gauge fields in the context of the Dirac equation
\cite{EPJC.2005.41.421,PRD.2013.87.047701}.
In the fermion sector, for example, this violation is implemented by
introducing two CPT-odd terms, $V_{\mu }\bar{\psi}\gamma^{\mu }\psi $,
$W_{\mu }\bar{\psi}\gamma _{5}\gamma ^{\mu }\psi $, where $V_{\mu }$,
$W_{\mu }$ are the LV backgrounds.

In this paper, we reexamine the Aharonov-Casher (AC) problem in a
CPT-odd Lorentz-violating background addressed in
Ref. \cite{JPG.2013.40.075007}.
We analyze the calculations of the derivation of the nonrelativistic
Hamiltonian, obtaining the correct Hamiltonian and supplying a condition
that establishes the exact equivalence between the original spin-1/2 AC
and LV-AC  effects.
We also solve the scattering and bound state problems of the model using
the physical regularization scheme
\cite{PRD.2012.85.041701,AoP.2013.339.510} based on the self-adjoint
extension method  proposed in
Refs. \cite{JMP.1985.26.2520,CMP.1991.139.103}.

The work is outlined in the following way:
In Section \ref{sec:motion}, we derive the equation of planar motion in
order to study the physical implications of the LV background on the
spin-1/2 AC problem.
We also obtain a condition that establishes the exact equivalence
between the original spin-1/2-AC and LV-AC effects.
The Section \ref{sec:selfae} is devoted to the study of the LV
Hamiltonian via the self-adjoint extension technique and are presented
some important properties of the LV wave function.
In Section \ref{sec:bound} the bound state energy is determined in terms
of the physics of the problem.
In Section \ref{sec:scattbound} are addressed the scattering and bound
state problems within the framework of the LV Schr\"{o}dinger-Pauli
equation.
Expressions for the bound state energies, phase-shift and scattering
matrix are computed and all them are explicitly described in terms of
the physical condition of the problem.
The self-adjoint extension parameter is also derived in terms of the
physical parameters.
At the end, we make a detailed analysis of the helicity conservation's
problem in the present framework.
In Section \ref{sec:conclusion} we give our conclusions and remarks.

\section{The equation of motion}
\label{sec:motion}

In this section, we derive the equation of motion that governs the
dynamics of a spin-1/2 neutral particle in a radial electric field and a
LV background vector.
We begin with the (3+1)-dimensional Dirac equation with a
LV and CPT-odd nonminimal coupling between fermions and the gauge field
as proposed in Ref. \cite{EPJC.2005.41.421} in the form 
($\hbar =c=1$ and signature $(+---)$ )
\begin{equation}
  \left[ i\gamma ^{\mu }D_{\mu }-M\right] \Psi =0,  \label{eq:diraccpt}
\end{equation}with
\begin{equation}
D_{\mu }=\partial _{\mu }+ieA_{\mu }+igV^{\nu }\tilde{F}_{\mu \nu },
\end{equation}were $\Psi $ is the fermion spinor of four-component,
$V^{\mu }=(\mathit{V}_{0},\mathbf{V})$ is the Carroll-Field-Jackiw
four-vector, $e$ is the eletric charge, $g$ is a constant that measures
the nonminimal coupling magnitude. 
The electromagnetic field tensor is given by
\begin{align}
  F^{\mu \nu}
  ={}&-F^{\nu \mu },  \nonumber \\
  ={}&\partial^{\mu }A^{\nu}-\partial ^{\nu }A^{\mu },  \nonumber \\
  ={}&\left(
       \begin{array}{cccc}
         0 & -E^{1} & -E^{2} & -E^{3} \\
         E^{1} & 0 & -B^{3} & B^{2} \\
         E^{2} & B^{3} & 0 & -B^{1} \\
         E^{3} & -B^{2} & B^{1} & 0
       \end{array}\right) ,  \label{eq:fmn}
\end{align}
where $A^{\mu }=\left( A^{0},\mathbf{A}\right) $ is the 4-vetor potential.
By using the Levi-Civita's antisymmetric symbol $\varepsilon ^{\mu \nu \rho
  \sigma }$ (with $\varepsilon _{\mu \nu \rho \sigma }=-\varepsilon ^{\mu \nu
\rho \sigma }$) to be equal to $1$ or $-1$ according to whether ($\mu \nu
\rho \sigma $) is an even or odd permutation of $(0,1,2,3)$ and zero
otherwise, we can obtain the dual tensor
\begin{equation}
  \tilde{F}^{\mu \nu }=-\tilde{F}^{\nu \mu }=
  \frac{1}{2}\varepsilon ^{\mu \nu\rho \sigma }F_{\rho \sigma }=
  \left(
    \begin{array}{cccc}
      0 & -B^{1} & -B^{2} & -B^{3} \\
      B^{1} & 0 & E^{3} & -E^{2} \\
      B^{2} & -E^{3} & 0 & E^{1} \\
      B^{3} & E^{2} & -E^{1} & 0
    \end{array}
  \right).  \label{eq:fdual}
\end{equation}
The tensor $\tilde{F}_{\mu \nu }$ is obtained directly from
$\tilde{F}^{\mu\nu }$ as 
\begin{align}
  \tilde{F}_{\mu \nu }
  ={}&g_{\mu \gamma }\tilde{F}^{\gamma\delta }g_{\delta\nu },  \nonumber \\
  ={}&\left(
       \begin{array}{cccc}
         0 & B^{1} & B^{2} & B^{3} \\
         -B^{1} & 0 & E^{3} & -E^{2} \\
         -B^{2} & -E^{3} & 0 & E^{1} \\
         -B^{3} & E^{2} & -E & 0
       \end{array}\right) .
\end{align}
From the above results, we identify
\begin{align}
\tilde{F}_{0i}={}& B^{i}=-B_{i}, \\
\tilde{F}_{ij}={}& \varepsilon ^{ijk}E^{k}=\varepsilon _{ijk}E_{k},
\end{align}so that the Dirac equation \eqref{eq:diraccpt} can be written
more explicitly as
\begin{align}
  \bar{\mathcal{E}}\psi
  = {} &
         \boldsymbol{\alpha}\cdot 
         \left[ \mathbf{p}-e\mathbf{A}-gV^{0}\mathbf{B}-g
         (\mathbf{V}\times \mathbf{E})\right] \psi
         \nonumber \\
       & +\left(eA_{0}+g\mathbf{V}\cdot \mathbf{B}+\beta M  \right) \psi,
         \label{eq:diracex}
\end{align}
where
\begin{equation}
  \beta=\gamma^{0}=
  \left(
    \begin{array}{cc}
      1 & 0 \\
      0 & -1
    \end{array}
  \right), \quad
  \boldsymbol{\gamma}=
  \left(
    \begin{array}{cc}
      0 & \boldsymbol{\sigma} \\
      -\boldsymbol{\sigma} & 0
    \end{array}
  \right) ,\quad
  \boldsymbol{\alpha}=
  \left(
    \begin{array}{cc}
      0 & \boldsymbol{\sigma} \\
      \boldsymbol{\sigma} & 0
    \end{array}
  \right) ,
\end{equation}
are the standard Dirac matrices and
$\boldsymbol{\sigma}=(\sigma _{1},\sigma_{2},\sigma _{3})$ are the Pauli
matrices.
Since we are only interested in planar dynamics of the AC problem, then
we specialize to the case $p_{3}=0$ in Eq. \eqref{eq:diracex} and consider
only the components $\tilde{F}_{ij}$ of $ \tilde{F}_{\mu \nu }$.
In this case, the relevant equation is the planar Dirac equation
\begin{equation}
\left\{ \mathbf{\alpha }\cdot \left[ \mathbf{p}-g\left( \mathbf{V}\times
\mathbf{E}\right) \right] +\beta M\right\} \psi =\mathcal{\bar{E}}\psi .
\label{eq:defdirac}
\end{equation}
Equation \eqref{eq:defdirac} in the nonrelativistic limit is found to be
\begin{equation}
\hat{H}\psi =\mathcal{E}\psi ,  \label{eq:dedfm}
\end{equation}
where
\begin{equation}
  \hat{H}=\frac{1}{2M}
  \left[\mathbf{p}-g(\mathbf{V}\times \mathbf{E})\right]
^{2}-\frac{1}{2M}gs\boldsymbol{\sigma}\cdot \left[ \mathbf{\nabla }\times
\left( \mathbf{V}\times \mathbf{E}\right) \right] ,  \label{eq:hdef}
\end{equation}
is the Hamiltonian operator.
The field configuration (in cylindrical coordinates) is given by
\begin{equation}
\mathbf{E}=\frac{2\lambda }{r}\mathbf{\hat{r}},~~~\boldsymbol{\nabla }\cdot
\mathbf{E}=2\lambda \frac{\delta (r)}{r},~~~V^{\mu }=(0,0,0,V_{z}),
\label{eq:fieldac}
\end{equation}
where $\mathbf{E}$, is the electric field generated by an infinite charged
filament and $\lambda $ is the charge density along the $z$-axis.

After substitution of Eq. \eqref{eq:fieldac} into \eqref{eq:hdef}, we find
\begin{equation}
  \hat{H}=\frac{1}{2M}
  \left[ \hat{H}_{0}-s\eta\sigma _{z}\frac{\delta (r)}{r}
\right] ,  \label{eq:hnrf}
\end{equation}
with
\begin{equation}
\hat{H}_{0}=\left( \frac{1}{i}\boldsymbol{\nabla}
  -\eta \frac{\hat{\boldsymbol{\varphi}}}{r}\right)^{2},
\label{eq:hzero}
\end{equation}
and
\begin{equation}
\eta =2\lambda gV_{z},  \label{eq:deltac}
\end{equation}
is the coupling constant of the $\delta(r)/r$ potential.

The Hamiltonian in Eq. \eqref{eq:hnrf} governs the quantum dynamics of a
spin-1/2 neutral particle with a radial electric field, i.e., a spin-1/2 AC
problem, with $g\mathbf{V}$ playing the role of a nontrivial magnetic dipole
moment. Also, note the presence of a $\delta $ function which is singular at
the origin in Eq. \eqref{eq:hnrf}. This makes the problem more complicated
to be solved. Such kind of point interaction potential can then be addressed
by the self-adjoint extension approach \cite
{CMP.1991.139.103,JMP.1985.26.2520,Book.2004.Albeverio}, which will be used
for studying the scattering and bound state scenarios.

Now, it is instructive to take look at the exact equivalence between the
LV-AC problem and the usual AC problem.
The Hamiltonian of the usual AC problem for a neutral particle with
magnetic moment $\boldsymbol{\mu}$ can be written as \cite{PRL.1990.64.2347}
\begin{equation}
  \hat{H}_{AC}=
  \left[
    \mathbf{p}+s\left( \boldsymbol{\mu}\times \mathbf{E}\right)
  \right]^{2}+
  \frac{1}{2M}\mu\sigma _{3}
  \left( \mathbf{\nabla }\cdot \mathbf{E}\right) ,
  \label{eq:hacu}
\end{equation}
which upon substitution of \eqref{eq:fieldac}, it assumes the form
\begin{equation}
  \hat{H}_{AC}=\frac{1}{2M}\left( \frac{1}{i}\boldsymbol{\nabla}
    +s\eta_{AC}\frac{\hat{\mathbf{\varphi }}}{r}\right) ^{2}
  +\frac{1}{2M}\eta_{AC}\sigma _{z}\frac{\delta (r)}{r},  \label{eq:hacu2}
\end{equation}
where
\begin{equation}
\eta_{AC}=2\lambda \mu .
\end{equation}
It is worth of mention that the operator in \eqref{eq:hacu2}
is formally the same as the two-dimensional spin-1/2 Aharonov-Bohm
Hamiltonian, with the delta function playing the role of the Zeeman
interaction between the spin and the magnetic flux tube
\cite{PRL.1990.64.503}.
This later problem has been solved in full details in
Refs. \cite{AoP.1983.146.1,LMP.1998.43.43,JMP.1998.39.47,PRL.1990.64.503}
(see also \cite{Book.2004.Albeverio} and references therein.)

We can observe in Eq. \eqref{eq:hacu} that the magnetic moment of the
particle, $\boldsymbol{\mu}=\mu \boldsymbol{\sigma}$, it is an spinor
quantity.
On the other hand, in the LV-AC Hamiltonian (Eq. \eqref{eq:hdef}), the
quantity $g\mathbf{V}$ has vectorial character.
In this way, the equivalence between the effects is achieved by
replacing
\begin{equation}
gV_{z} \to -\mu s,
\end{equation}
directly in Eq. \eqref{eq:hnrf}.
Thus, we have established an exact equivalence between the usual AC and
LV-AC effects.

\section{Self-adjoint extension analysis}
\label{sec:selfae}

An operator $\mathcal{O}$, with domain
$\mathcal{D}(\mathcal{O})$, is said to be self-adjoint if and
only if
$\mathcal{D}(\mathcal{O}^{\dagger})=\mathcal{D}(\mathcal{O})$
and $\mathcal{O}^{\dagger}=\mathcal{O}$.
In order to determine all self-adjoint extensions of
\eqref{eq:hzero}, making use of the underlying rotational
symmetry expressed by the fact that $[\hat{H},\hat{J}_{z}]=0$,
where $\hat{J}_{z}=-i\partial/\partial_{\varphi}+\sigma_{z}/2$ is
the total angular momentum operator in the $z$-direction.
Furthermore, the Hilbert space
$\mathfrak{H}=L^{2}(\mathbb{R}^{2})$ is decomposed
with respect to the total angular momentum
$\mathfrak{H}=\mathfrak{H}_{r}\otimes\mathfrak{H}_{\varphi}$,
where $\mathfrak{H}_{r}=L^{2}(\mathbb{R}^{+},rdr)$ and
$\mathfrak{H}_{\varphi}=L^{2}(\mathcal{S}^{1},d\varphi)$,
with $\mathcal{S}^{1}$ denoting the unit sphere in
$\mathbb{R}^{2}$.
So, it is possible to express the eigenfunctions of the two
dimensional Hamiltonian in terms of the eigenfunctions of
$\hat{J}_{z}$
\begin{equation}
  \Psi(r,\varphi)=
  \left(
    \begin{array}{c}
      \psi_{m}(r) e^{i(m_{j}-1/2)\varphi } \\
      \chi_{m}(r) e^{i(m_{j}+1/2)\varphi }
    \end{array}
  \right),
\label{eq:wavef}
\end{equation}
with $m_{j}=m+1/2=\pm 1/2,\pm 3/2,\ldots $, and $m\in\mathbb{Z}$.
By inserting Eq. \eqref{eq:wavef} into Eq. \eqref{eq:dedfm} the
Schr\"{o}dinger-Pauli equation for $\psi_{m}(r)$ is found to be
($k^{2}=2ME$)
\begin{equation}
  H\psi_{m}(r)=k^{2}\psi_{m}(r),
  \label{eq:eigen}
\end{equation}
where
\begin{equation}
  H=H_{0}-s\eta \frac{\delta(r)}{r},
  \label{eq:hfull}
\end{equation}
and
\begin{equation}
  H_{0}=
  -\frac{d^{2}}{dr^{2}}-\frac{1}{r}\frac{d}{dr}
  +\frac{(m-\eta)^{2}}{r^{2}}.
\end{equation}

The self-adjoint extension approach consists, essentially, in
extending the domain $\mathcal{D}(H_{0})$ to match
$\mathcal{D}(H_{0}^{\dagger})$ and therefore turning $H_{0}$
into a self-adjoint operator.
To do so, we must find the deficiency subspaces,
$N_{\pm}$, with dimensions $n_{\pm}$, which are called
deficiency indices of $H_{0}$ \cite{Book.1975.Reed.II}.
A necessary and sufficient condition for $H_{0}$ being
essentially self-adjoint is that $n_{+}=n_{-}=0$.
On the other hand, if $n_{+}=n_{-}\geq 1$, then $H_{0}$ will
have an infinite number of self-adjoint extensions parametrized
by unitary operators  $U_{\theta}:N_{+}\to N_{-}$, with  $\theta\in
[0,2\pi)$.
In order to find the deficiency subspaces of $H_{0}$ in
$\mathfrak{H}_{r}$, we must solve the eigenvalue equation
\begin{equation}
  H_{0}^{\dagger}\psi_{\pm}
  =\pm i k_{0}^{2} \psi_{\pm},
  \label{eq:eigendefs}
\end{equation}
where $k_{0}^{2}\in\mathbb{R}$ was introduced for dimensional
reasons.
Since $H_{0}^{\dagger}=H_{0}$, the solutions of
Eq. \eqref{eq:eigendefs} which vanishes at the infinite are the
modified Bessel functions of second kind (up to a constant)
\begin{equation}
  \psi_{\pm}=K_{|m-\eta|}(\varepsilon_{\pm} r),
  \label{eq:psidefs}
\end{equation}
with $\varepsilon_{\pm}=e^{\mp i \pi/4} k_{0}$.
The solutions $\psi_{\pm}$ are normalizable if and only if $|m-\eta|<1$.  
The dimension of such deficiency subspace is thus
$(n_{+},n_{-})=(1,1)$.
According to the von Neumann-Krein theory, all self-adjoint
extensions $H_{\theta,0}$ of $H_{0}$ are given by the
one-parameter family
\begin{equation}
\mathcal{D}(H_{\theta,0})=
\mathcal{D}(H_{0})\oplus (I +U_{\theta})N_{+}.
\end{equation}
Thus, $\mathcal{D}(H_{\theta,0})$ in $\mathfrak{H}_{r}$ is given
by the set of functions \cite{Book.1975.Reed.II}
\begin{equation}
  \label{eq:domain}
  \psi_{\theta}(r) =  \psi_{m}(r)+
  c\left[
    K_{|m-\eta|}(\varepsilon_{+} r)+
    e^{i\theta}K_{|m-\eta|}(\varepsilon_{-} r)
  \right],
\end{equation}
where $\psi_{m}(r)$, with $\psi_{m}(0)=\dot{\psi}_{m}(0)=0$
($\dot{\psi}\equiv d\psi/dr$), is a regular wave function,
$c\in \mathbb{C}$ and the number $\theta \in [0,2\pi)$
represents a choice for the boundary condition.
Using the unitary operator
$U: L^{2}(\mathbb{R}^{+},rdr) \to L^{2}(\mathbb{R}^{+}, dr)$,
given by $(U \xi)(r)=r^{1/2}\xi(r)$, the operator $H_{0}$
reads
\begin{equation}
  \tilde{H}_{0}=
  U H_{0} U^{-1}=
  -\frac{d^{2}}{dr^{2}}
  -\frac{(m-\eta)^{2}-1/4}{r^{2}}.
\end{equation}
By standard results, the radial operator $\tilde{H}_{0}$
is essentially self-adjoint for $|m-\eta| \geq 1$, while
for $|m-\eta|< 1$ it admits an one-parameter family of
self-adjoint extensions \cite{Book.1975.Reed.II}.
This statement can be understood based in
Eq. \eqref{eq:psidefs}, because for $|m-\eta| \geq 1$ the
right hand side is not in $\mathfrak{H}_{r}$ at $0$, while it is
in  $\mathfrak{H}_{r}$ for $|m-\eta| < 1$.
To characterize the one-parameter family of the self-adjoint
extension, we will use the KS \cite{CMP.1991.139.103} and BG
\cite{JMP.1985.26.2520} approaches, both base in boundary
condition at the origin, as we explain below.

\subsection{KS self-adjoint extension approach}
\label{sec:KS-approach}

Following the Ref. \cite{CMP.1991.139.103}, in the KS approach,
the boundary condition is a match of the logarithmic derivatives
of the zero-energy solutions for the problem $H_{0}$ plus self-adjoint
extension, i.e., one considers the zero-energy solutions
$\psi_{0}$ and $\psi_{\theta,0}$ for $H$ and $H_{0}$,
respectively,
\begin{equation}
  \left[
    \frac{d^{2}}{dr^{2}}+\frac{1}{r}\frac{d}{dr}-
    \frac{(m-\eta)^{2}}{r^{2}}+ s\eta\frac{\delta(r)}{r}
\right] \psi_{0}=0,
  \label{eq:statictrue}
\end{equation}
and
\begin{equation}
  \left[
    \frac{d^{2}}{dr^{2}}+\frac{1}{r}\frac{d}{dr}-
    \frac{(m-\eta)^{2}}{r^{2}}
  \right] \psi_{\theta,0}=0.
  \label{eq:thetastatic}
\end{equation}
A fitting value for the number $\theta$ is determined by the
boundary condition at the origin
\begin{equation}
  a\frac{\dot{\psi}_{0}}{\psi_{0}}\Big|_{r=a}=
  a\frac{\dot{\psi}_{\theta,0}}{\psi_{\theta,0}}\Big|_{r=a}.
\label{eq:logder}
\end{equation}
where $a$ is a very small radius, being smaller than the Compton wave
length $\lambda_{C}$ of the electron \cite{PLB.1994.333.238},
which comes form the regularization of the $\delta$ function
\cite{PRD.2012.85.041701,AoP.2013.339.510}.
The present approach has the advantage of yielding the number
$\theta$ in terms of the physics of the problem, but is only
applicable for determination of bound states, being not
appropriate for dealing with scattering problems.

\subsection{BG self-adjoint extension approach}
\label{sec:BG-approach}

As mentioned above, the KS approach is suitable to address only
bound states.
On the other hand, the BG method is suitable to address both
bound and scattering scenarios, with the disadvantage of
allowing arbitrary self-adjoint extension parameter.
By comparing the results of these two approaches for bound
states, the self-adjoint extension parameter can be determined
in terms of the physics of the problem.

In the BG approach, the boundary condition is a mathematical
limit allowing divergent solutions of the Hamiltonian $H_{0}$ at
isolated points, provided they remain square integrable.
All the self-adjoint extensions $H_{0,\lambda_{m}}$ of
$\tilde{H}_{0}$ are parametrized by the boundary condition at
the origin \cite{JMP.1985.26.2520,Book.2004.Albeverio}
\begin{equation}
\psi^{(0)}=\lambda_{m}\psi^{(1)},  \label{eq:bc}
\end{equation}
with
\begin{align*}
  \psi^{(0)}={} & \lim_{r\rightarrow 0^{+}}r^{|m-\eta|}\psi(r), \\
  \psi^{(1)}={} & \lim_{r\rightarrow 0^{+}}\frac{1}{r^{|m-\eta|}}
  \left[\psi(r)-\psi^{(0)}\frac{1}{r^{|m-\eta|}}\right],
\end{align*}
where $\lambda_{m}$ is the self-adjoint extension parameter.
In \cite {Book.2004.Albeverio} is shown that there is a
relation between the self-adjoint extension parameter
$\lambda_{m}$ and the number $\theta$ in the KS approach.
The number $\theta$ is associated with the mapping of
deficiency subspaces and extend the domain of operator to make
it self-adjoint.
The self-adjoint extension parameter $\lambda_{m}$ have a
physical interpretation: it represents the scattering length
\cite{Book.2011.Sakurai} of $H_{0,\lambda_{m}}$
\cite{Book.2004.Albeverio}.
For $\lambda_{m}=0$, we have the free Hamiltonian (without the
$\delta$ function) with regular wave functions at origin and
for $\lambda_{m}\neq 0$ the boundary condition in \eqref{eq:bc}
allows a $r^{-|m-\eta|}$ singularity in the wave functions at
origin.

\section{Bound state analysis}
\label{sec:bound}

In this section we employ the KS approach for determination of
the bound states for the Hamiltonian in $H$.
Thus, the first term of Eq. \eqref{eq:logder} is obtained by
integrating the Eq. \eqref{eq:statictrue} from 0 to $a$.
The second term is calculated using the asymptotic
representation for the Bessel function $K_{|m-\eta|}$ for small
argument.
So, from \eqref{eq:logder} we arrive at
\begin{equation}
  a\dot{\Upsilon}_{\theta}(a)+s\eta \Upsilon_{\theta}=0,
  \label{eq:saepapprox}
\end{equation}
with
\begin{equation}
\Upsilon_{\theta}(r)=D(\varepsilon_{+})+e^{i\theta} D(\varepsilon_{-}) ,
\end{equation}
and
\begin{equation}
  D(\varepsilon_{\pm}) =
  \frac
  {\left( \varepsilon_{\pm} r\right)^{-|m-\eta|}}
  {2^{-|m-\eta|}\Gamma(1-|m-\eta|)}
  -\frac
  {\left( \varepsilon_{\pm} r\right)^{|m-\eta|}}
  {2^{|m-\eta|}\Gamma(1+|m-\eta|)}.
\end{equation}
Eq. \eqref{eq:saepapprox} gives us the parameter $\theta$ in
terms of the physics of the problem, i.e., the correct behavior
of the wave functions at the origin.

Next, we will find the bound states of the Hamiltonian $H_{0}$
and, by using \eqref{eq:saepapprox}, the spectrum of
$H$ will be determined without any arbitrary parameter.
Then, from $H_{0}\psi_{\theta}=E_{b}\psi_{\theta}$ we
achieve the modified Bessel equation
($\kappa^{2}=-2ME_{b}$)
\begin{equation}
  \left[
    \frac{d^{2}}{dr^{2}}+\frac{1}{r}\frac{d}{dr}-
    \frac{(m-\eta)^{2}}{r^{2}}-\kappa ^{2}
  \right]\psi_{\theta}(r)=0,
  \label{eq:eigenvalue}
\end{equation}
where $E_{b}<0$  (since we are looking for bound states).
The general solution for the above equation is
\begin{equation}
\psi_{\theta}(r)=K_{|m-\eta|}\left(r\sqrt{-2ME_{b}}\right).
\label{eq:sver}
\end{equation}
Since these solutions belong to $\mathcal{D}(H_{\theta,0})$,
they present the form \eqref{eq:domain} for a
$\theta$ selected from
the physics of the problem (cf. Eq. \eqref{eq:saepapprox}).
So, we substitute \eqref{eq:sver} into \eqref{eq:domain} and
compute $a{\dot{\psi}_{\theta}}/{\psi_{\theta}}|_{r=a}$.
After a straightforward calculation, we have the relation
\begin{equation}
  \frac
  {|m-\eta|
    \left[
      a^{2|m-\eta|}(-ME_{b})^{|m-\eta|}\Theta-1
    \right]
  }
  {a^{2|m-\eta|}(-ME_{b})^{|m-\eta|}\Theta+1}
    =-s\eta,
\end{equation}
where $\Theta=\Gamma(-|m-\eta|)/(2^{|m-\eta|}\Gamma(|m-\eta|))$.
Solving the above equation for $E_{b}$, we find the sought
energy spectrum
\begin{equation}
  E_{b}=
  -\frac{2}{M a^{2}}
  \left[
    \left(
      \frac
      {s\eta - |m-\eta|}
      {s\eta + |m-\eta|}
    \right)
    \frac
    {\Gamma (1+|m-\eta|)}
    {\Gamma (1-|m-\eta|)}
  \right]^{{1}/{|m-\eta|}}.
\label{eq:energy_KS}
\end{equation}
In the above relation, to ensure that the energy is a real
number, we must have $|s\eta| \geq |m-\eta|$, and due to
$|m-\eta|<1$ it is sufficient to consider $|s\eta|\geq 1$.
A necessary condition for a $\delta$ function generating an
attractive potential, able to support bound states,
is that the coupling constant must be negative.
Thus, the existence of bound states with real energies requires
\begin{equation}
s\eta \geq 1.
\end{equation}
From the above equation and Eq \eqref{eq:deltac}  it
follows that $s\lambda g V_{z} $ must be positive,
and consequently, there is a minimum value for this product.

\section{Scattering analysis}
\label{sec:scattbound}

In this section, we are interested in a situation in which a
incident particle reaches the center at $r=0$ and is scattered
out by the potential $-s\eta \delta(r)/r$.
The phase-shift, scattering amplitude and so on, are obtained by
employing the BG approach.
The equation to be solved is the Eq. \eqref{eq:eigen} and its
solution in the $r\neq 0$ region can be written as
\begin{equation}
\label{eq:sol1}
\psi_{m}(r)=a_{m}J_{|m-\eta|}(kr)+b_{m}Y_{|m-\eta|}(kr),
\end{equation}
with $a_{m}$ and $b_{m}$ being constants and $J_{\nu}(z)$ and
$Y_{\nu}(z)$ are the Bessel functions of first and second kind,
respectively.
Upon replacing $\psi_{m}(r)$ in the boundary condition
\eqref{eq:bc}, one obtain
\begin{align}
  \lambda_{m} a_{m} \mathcal{A} k^{|m-\eta|}= {}
  &
  b_{m}\bigg(\mathcal{B} k^{-|m-\eta|}
  - \lambda_{m}\mathcal{C}  k^{|m-\eta|}\\
  & - \lambda_{m}\mathcal{B} \mathcal{D} k^{-|m-\eta|}
  \lim_{r\rightarrow 0^{+}}r^{2-2|m-\eta|}\bigg),
  \label{eq:bcf}
\end{align}
with
\begin{align}
\mathcal{A} = {} & \frac{1}{2^{|m-\eta|}\Gamma(1+|m-\eta|)}, \nonumber \\
\mathcal{B} = {} & -\frac{2^{|m-\eta|}\Gamma(|m-\eta|)}{\pi},
\nonumber \\
\mathcal{C} = {} & -\frac{\cos (\pi |m-\eta|)
 \Gamma(-|m-\eta|)}{\pi 2^{|m-\eta|}}, \nonumber \\
\mathcal{D} = {} & \frac{k^{2}}{4(1-|m-\eta|)}.
\end{align}
In Eq. \eqref{eq:bcf}, $\lim_{r\rightarrow 0^{+}}r^{2-2|m-\eta|}$
is divergent if $|m-\eta|\geq 1$, hence $b_{m}$ must be zero.
On the other hand, $\lim_{r\rightarrow 0^{+}}r^{2-2|m-\eta|}$
is finite for $|m-\eta|<1$.
This means that there arises the contribution of the irregular
solution $Y_{|m-\eta|}(kr)$.
Here, the presence of an irregular solution contributing to the
wave function stems from the fact the Hamiltonian $H_{0}$ is not a
self-adjoint operator when $|m-\eta|<1$ (cf., Section
\ref{sec:selfae}).
Hence such irregular solution must be
associated with a self-adjoint extension of the operator $H_{0}$
\cite{JPA.1995.28.2359,PRA.1992.46.6052}.
Thus, for $|m-\eta|<1$, we have
\begin{equation}
  \lambda_{m} a_{m}\mathcal{A} k^{|m-\eta|}=
  b_{m}(\mathcal{B} k^{-|m-\eta|}-
  \lambda_{m}\mathcal{C} k^{|m-\eta|}),
\end{equation}
and by substituting the values of $\mathcal{A}$, $\mathcal{B}$
and $\mathcal{C}$ into above expression we find
\begin{equation}
  b_{m}=-\mu_{m}^{\lambda_{m}}(k,\eta) a_{m},
\end{equation}
where
\begin{equation}
  \mu_{m}^{\lambda_{m}}(k,\eta)=
  \frac
  {\lambda_{m} k^{2|m-\eta|}
    \Gamma(1-|m-\eta|)\sin(\pi |m-\eta|)}
  {B_{k}},
  \label{eq:mul}
\end{equation}
and
\begin{align}
  B_{k} = {} &
  \lambda_{m} k^{2|m-\eta|}
  \Gamma{(1-|m-\eta|)}\cos(\pi|m-\eta|)
  \nonumber\\
  &
  + 4^{|m-\eta|}\Gamma(1+|m-\eta|).
\end{align}
Since a $\delta$ function is a very short range potential, it
follows that the asymptotic behavior of $\psi_{m}(r)$ for
$r\rightarrow \infty$ is given by \cite{JPA.2010.43.354011}
\begin{equation}
  \psi_{m}(r)\sim \sqrt{\frac{2}{\pi kr}}
  \cos \left( kr-\frac{|m|\pi}{2}-
  \frac{\pi}{4}+
  \delta_{m}^{{\lambda_{m}}}(k,\eta)\right) ,
\label{eq:f1asim}
\end{equation}
where $\delta_{m}^{{\lambda_{m}}}(k,\eta)$  is a
scattering phase shift.
The phase shift is a measure of the argument difference to the
asymptotic behavior of the solution $J_{|m|}(kr)$ of the radial
free equation which is regular at the origin.
By using the asymptotic behavior of the Bessel functions
\cite{Book.1972.Abramowitz} into Eq. \eqref{eq:sol1} one obtain
\begin{align}
  \label{eq:scattsol}
  \psi_{m}(r)
   \sim
  &
  \sqrt{\frac{2}{\pi kr}}
  \left[
    \cos\left(kr-\frac{\pi |m-\eta|}{2}-\frac{\pi}{4}\right)
  \right. \nonumber \\
  &
  \left.
    - \mu_{m}^{{\lambda_{m}}}(k,\eta)
    \sin \left( kr-\frac{\pi |m-\eta|}{2}-\frac{\pi}{4}\right)
  \right] .
\end{align}
By comparing the above expression with Eq. \eqref{eq:f1asim}, we
have
\begin{equation}
\delta_{m}^{{\lambda_{m}}}(k,\eta)=
\Delta_{m}(\eta)+\theta_{m}^{\lambda_{m}}(k,\eta),
\label{eq:phaseshift}
\end{equation}
with
\begin{equation}
\Delta_{m}(\eta)=\frac{\pi}{2}(|m|-|m-\eta|),
\end{equation}
the phase shift of the AC scattering and
\begin{equation}
  \theta_{m}^{\lambda_{m}}(k,\eta)=\arctan {[\mu_{m}^{\lambda_{m}}(k,\eta)]}.
\end{equation}
Therefore, the scattering operator $S_{m}^{\lambda_{m}}(k,\eta)$
($S$-matrix) for the self-adjoint extension is
\begin{equation}
S_{m}^{\lambda_{m}}(k,\eta)=
e^{2i\delta_{m}^{{\lambda_{m}}}(k,\eta)}=
  \left[
    \frac
    {1+i\mu_{m}^{\lambda_{m}}(k,\eta)}
    {1-i\mu_{m}^{\lambda_{m}}(k,\eta)}
  \right]
  e^{2i\Delta_{m}(\eta)}.
\end{equation}
Using Eq. \eqref{eq:mul}, we have
\begin{equation}
  S_{m}^{\lambda_{m}}(k,\eta)
  = e^{2i\Delta_{m}(\eta)}
  \frac{\Omega_{+}}{\Omega_{-}},
  \label{eq:smatrix}
\end{equation}
with
\begin{equation}
\Omega_{\pm} = B_{k} \pm i \lambda_{m} k^{2|m-\eta|}
      \Gamma(1-|m-\eta|)\sin{(i\pi |m-\eta|)}.
\end{equation}
Hence, for any value of the self-adjoint extension parameter
$\lambda_{m}$, there is an additional scattering.
If ${\lambda_{m}}=0$, we achieve the corresponding result
for the AC problem with Dirichlet boundary condition
\cite{AoP.1983.146.1}, i.e.,
$S_{m}^{0}(k,\eta)=e^{2i\Delta_{m}(\eta)}$.
If we make ${\lambda_{m}}=\infty$, we get
$S_{m}^{\infty}(k,\eta)=e^{2i\Delta_{m}(\eta)+2i\pi |m-\eta|}$.

In accordance with the general theory of scattering, the poles
of the $S$-matrix in the upper half of the complex plane
\cite{PRC.1999.60.34308} determine the positions of the bound
states in the energy scale.
These poles occur in the denominator of \eqref{eq:smatrix} with
the replacement $k\rightarrow i\kappa$.
Thus,
\begin{equation}
  \Omega_{-}=0.
\end{equation}
Solving the above equation for $E_{b}$, we found the bound state
energy
\begin{equation}
  E_{b}=-\frac{2}{M}
  \left[-\frac{1}{\lambda_{m}}
    \frac{\Gamma(1+|m-\eta|)}{\Gamma(1-|m-\eta|)}
  \right]^{1/|m-\eta|},
  \label{eq:energy_BG-sc}
\end{equation}
for $\lambda_{m}<0$.
Hence, the poles of the scattering matrix only occur for
negative values of the self-adjoint extension parameter, when we
have scattering and bound states.
In this latter case, the scattering operator can be expressed in
terms of the bound state energy
\begin{equation}
  S_{m}^{\lambda_{m}}(k,\eta)=e^{2i\Delta_{m}(\eta)}
  \left[
    \frac
    {e^{2 i \pi |m-\eta|}-(\kappa/k)^{2|m-\eta|}}
    {1-(\kappa/k)^{2|m-\eta|}}
  \right].
\end{equation}

By comparing Eq. \eqref {eq:energy_BG-sc} with the
Eq. \eqref{eq:energy_KS} we have
\begin{equation}
  \frac{1}{\lambda_{m}}=
  -\frac{1}{a^{2|m-\eta|}}
    \left(
      \frac
      {s\eta - |m-\eta|}
      {s\eta + |m-\eta|}
    \right).
\label{eq:saep}
\end{equation}
We have thus attained a relation between the self-adjoint
extension parameter and the physical parameters of the problem.
It should be mentioned that some relations involving the
self-adjoint extension parameter and the $\delta$-function
coupling constant were previously obtained by using Green's
function in Ref. \cite{JMP.1995.36.5453} and the renormalization
technique in Ref. \cite{Book.1995.Jackiw}, being both, however,
deprived from a clear physical interpretation.

The scattering amplitude $f(k,\eta)$ can be now obtained using
the standard methods of scattering theory, namely
 ($f_{k}=1/\sqrt{2\pi i k}$)
\begin{align}
  f(k,\eta)
  = {} &
  f_{k}\sum_{m=-\infty}^{\infty}
  \left(
    e^{2 i\delta_{m}^{{\lambda_{m}}}(k,\eta)}-1
  \right)
  e^{im\varphi}   \nonumber \\
  ={} &
  f_{k} \Bigg \{
  \sum_{|m-\eta|\geq 1}
  (e^{2 i \Delta_{m}(\eta)}-1)e^{im\varphi}
  \nonumber \\
  & +
   \sum_{|m-\eta|< 1}
  \left\{
    e^{2 i \Delta_{m}(\eta)}
    \left[
      \frac
      {1+ i\mu_{m}^{\lambda_{m}}(k,\eta)}
      {1-i\mu_{m}^{\lambda_{m}}(k,\eta)}
    \right]-1
  \right\}
    \nonumber \\
  &\times e^{im\varphi}\Bigg\}.
  \label{eq:scattamp}
\end{align}
The first sum is the AC amplitude (i.e., in the absence of
the $\delta$ function), while the second sum is the
contribution that come from the singular solutions.
In the above equation we can see that the scattering amplitude
is energy dependent (cf., Eq. \eqref{eq:mul}).
This is a clearly manifestation of the known non-conservation of
the helicity in the AC scattering \cite{PRL.1990.64.2347},
because the only length scale in the nonrelativistic problem is
set by  $1/k$, so it follows that the scattering amplitude would
be a function of the angle alone, multiplied by $1/k$
\cite{PRD.1977.16.1815}.
In fact, the failure of helicity conservation expressed in
Eq. \eqref{eq:scattamp}, it stems from the fact that the
$\delta$ function singularity make the Hamiltonian and the
helicity nonself-adjoint operators
\cite{PRD.1977.15.2287,PRL.1983.50.464,PLB.1993.298.63,
NPB.1994.419.323}.
By expressing the helicity operator,
$\hat{h}=\boldsymbol{\Sigma} \cdot \boldsymbol{\Pi}$,
in terms of the variables used in \eqref{eq:wavef}, we attain
\begin{equation}
  \hat{h} =
  \left(
    \begin{array}{cc}
      0
      & \displaystyle -i
      \left(
        \partial_{r}+\frac{|m-\eta|+1}{r}
      \right) \\
      \displaystyle -i
      \left(
        \partial_{r}-\frac{|m-\eta|}{r}
      \right)
      & 0
    \end{array}
  \right).
\end{equation}
Notice that under a parity $\pi$ transformation
$\hat{h}\to\pi^{\dagger}\hat{h}\pi=-\hat{h}$, that comes
immediately from the parity transformation
$\pi^{\dagger}r\pi=-r$.
This is in fact the helicity odd-parity property.
The helicity operator share the same issue as the Hamiltonian
operator in the interval $|m-\eta|<1$, i.e., it is not
self-adjoint \cite{PRD.1994.49.2092,JPA.2001.34.8859}.
Despite that on a finite interval $[0,L]$, $\hat{h}$  is a
self-adjoint operator with domain in the functions satisfying
$\xi(L)=e^{i\theta}\xi{(0)}$, it does not admit a self-adjoint
extension on the interval $[0,\infty)$ \cite{AJP.2001.69.322},
and consequently it can be not conserved, thus the helicity
conservation is broken due to the presence of the singularity at
the origin \cite{PRD.1977.16.1815,PRL.1983.50.464}.

\section{Conclusion}
\label{sec:conclusion}

We have studied the spin-1/2 AC bound and scattering problem
with a Lorentz-violating and CPT-odd nonminimal coupling between
fermions and the gauge field in the context of the Dirac
equation.
The self-adjoint extension approach was used to determine the
bound states of the  particle in terms of the physics of the
problem, in a very consistent way and without any arbitrary
parameter.
It has been shown that there is an additional scattering for any
value of the self-adjoint extension parameter and for negative
values of this parameter there is bound states.
By comparing the results from bound and scattering scenarios,
the self-adjoint extension parameter was determined.
The scattering amplitude show a energy dependency, so the
helicity in not conserved.
This stem from the fact that the helicity operator is not
a self-adjoint extension operator.
Therefore, it does not represent a quantum observable and does
not correspond to a conserved quantity.

\section*{Acknowledgments}
We would like to thank R. Casana and M. M. Ferreira Jr. for fruitful
discussions.
We also would like to thank the anonymous referee for the useful
comments and suggestions.
This work was supported by the 
Conselho Nacional de Desenvolvimento Cient\'{\i}fico e Tecnol\'{o}gico
(Grants No. 482015/2013-6 (Universal), No. 306068/2013-3 (PQ), No
460404/2014-8 (Universal) and 206224/2014-1 (PDE)) and 
FAPEMA (Grant No. 00845/13).

\section*{References}

\bibliography{bibliography}

\begin{thebibliography}{75}%
\makeatletter
\providecommand \@ifxundefined [1]{%
 \@ifx{#1\undefined}
}%
\providecommand \@ifnum [1]{%
 \ifnum #1\expandafter \@firstoftwo
 \else \expandafter \@secondoftwo
 \fi
}%
\providecommand \@ifx [1]{%
 \ifx #1\expandafter \@firstoftwo
 \else \expandafter \@secondoftwo
 \fi
}%
\providecommand \natexlab [1]{#1}%
\providecommand \enquote  [1]{``#1''}%
\providecommand \bibnamefont  [1]{#1}%
\providecommand \bibfnamefont [1]{#1}%
\providecommand \citenamefont [1]{#1}%
\providecommand \href@noop [0]{\@secondoftwo}%
\providecommand \href [0]{\begingroup \@sanitize@url \@href}%
\providecommand \@href[1]{\@@startlink{#1}\@@href}%
\providecommand \@@href[1]{\endgroup#1\@@endlink}%
\providecommand \@sanitize@url [0]{\catcode `\\12\catcode `\$12\catcode
  `\&12\catcode `\#12\catcode `\^12\catcode `\_12\catcode `\%12\relax}%
\providecommand \@@startlink[1]{}%
\providecommand \@@endlink[0]{}%
\providecommand \url  [0]{\begingroup\@sanitize@url \@url }%
\providecommand \@url [1]{\endgroup\@href {#1}{\urlprefix }}%
\providecommand \urlprefix  [0]{URL }%
\providecommand \Eprint [0]{\href }%
\providecommand \doibase [0]{http://dx.doi.org/}%
\providecommand \selectlanguage [0]{\@gobble}%
\providecommand \bibinfo  [0]{\@secondoftwo}%
\providecommand \bibfield  [0]{\@secondoftwo}%
\providecommand \translation [1]{[#1]}%
\providecommand \BibitemOpen [0]{}%
\providecommand \bibitemStop [0]{}%
\providecommand \bibitemNoStop [0]{.\EOS\space}%
\providecommand \EOS [0]{\spacefactor3000\relax}%
\providecommand \BibitemShut  [1]{\csname bibitem#1\endcsname}%
\let\auto@bib@innerbib\@empty
\bibitem [{\citenamefont {Colladay}\ and\ \citenamefont
  {Kosteleck\'y}(1997)}]{PRD.1997.55.6760}%
  \BibitemOpen
  \bibfield  {author} {\bibinfo {author} {\bibfnamefont {D.}~\bibnamefont
  {Colladay}}\ and\ \bibinfo {author} {\bibfnamefont {V.~A.}\ \bibnamefont
  {Kosteleck\'y}},\ }\href {\doibase 10.1103/PhysRevD.55.6760} {\bibfield
  {journal} {\bibinfo  {journal} {Phys. Rev. D}\ }\textbf {\bibinfo {volume}
  {55}},\ \bibinfo {pages} {6760} (\bibinfo {year} {1997})}\BibitemShut
  {NoStop}%
\bibitem [{\citenamefont {Colladay}\ and\ \citenamefont
  {Kosteleck\'y}(1998)}]{PRD.1998.58.116002}%
  \BibitemOpen
  \bibfield  {author} {\bibinfo {author} {\bibfnamefont {D.}~\bibnamefont
  {Colladay}}\ and\ \bibinfo {author} {\bibfnamefont {V.~A.}\ \bibnamefont
  {Kosteleck\'y}},\ }\href {\doibase 10.1103/PhysRevD.58.116002} {\bibfield
  {journal} {\bibinfo  {journal} {Phys. Rev. D}\ }\textbf {\bibinfo {volume}
  {58}},\ \bibinfo {pages} {116002} (\bibinfo {year} {1998})}\BibitemShut
  {NoStop}%
\bibitem [{\citenamefont {Kosteleck\'y}(2004)}]{PRD.2004.69.105009}%
  \BibitemOpen
  \bibfield  {author} {\bibinfo {author} {\bibfnamefont {V.~A.}\ \bibnamefont
  {Kosteleck\'y}},\ }\href {\doibase 10.1103/PhysRevD.69.105009} {\bibfield
  {journal} {\bibinfo  {journal} {Phys. Rev. D}\ }\textbf {\bibinfo {volume}
  {69}},\ \bibinfo {pages} {105009} (\bibinfo {year} {2004})}\BibitemShut
  {NoStop}%
\bibitem [{\citenamefont {Cambiaso}\ \emph {et~al.}(2014)\citenamefont
  {Cambiaso}, \citenamefont {Lehnert},\ and\ \citenamefont
  {Potting}}]{PRD.2014.90.065003}%
  \BibitemOpen
  \bibfield  {author} {\bibinfo {author} {\bibfnamefont {M.}~\bibnamefont
  {Cambiaso}}, \bibinfo {author} {\bibfnamefont {R.}~\bibnamefont {Lehnert}}, \
  and\ \bibinfo {author} {\bibfnamefont {R.}~\bibnamefont {Potting}},\ }\href
  {\doibase 10.1103/PhysRevD.90.065003} {\bibfield  {journal} {\bibinfo
  {journal} {Phys. Rev. D}\ }\textbf {\bibinfo {volume} {90}},\ \bibinfo
  {pages} {065003} (\bibinfo {year} {2014})}\BibitemShut {NoStop}%
\bibitem [{\citenamefont {Schreck}(2014)}]{PRD.2014.89.105019}%
  \BibitemOpen
  \bibfield  {author} {\bibinfo {author} {\bibfnamefont {M.}~\bibnamefont
  {Schreck}},\ }\href {\doibase 10.1103/PhysRevD.89.105019} {\bibfield
  {journal} {\bibinfo  {journal} {Phys. Rev. D}\ }\textbf {\bibinfo {volume}
  {89}},\ \bibinfo {pages} {105019} (\bibinfo {year} {2014})}\BibitemShut
  {NoStop}%
\bibitem [{\citenamefont {Villalobos}\ \emph {et~al.}(2014)\citenamefont
  {Villalobos}, \citenamefont {da~Silva}, \citenamefont {Hott},\ and\
  \citenamefont {Belich}}]{EPJC.2014.74.2799}%
  \BibitemOpen
  \bibfield  {author} {\bibinfo {author} {\bibfnamefont {C.}~\bibnamefont
  {Villalobos}}, \bibinfo {author} {\bibfnamefont {J.}~\bibnamefont
  {da~Silva}}, \bibinfo {author} {\bibfnamefont {M.}~\bibnamefont {Hott}}, \
  and\ \bibinfo {author} {\bibfnamefont {H.}~\bibnamefont {Belich}},\ }\href
  {\doibase 10.1140/epjc/s10052-014-2799-1} {\bibfield  {journal} {\bibinfo
  {journal} {Eur. Phys. J. C}\ }\textbf {\bibinfo {volume} {74}},\ \bibinfo
  {pages} {2799} (\bibinfo {year} {2014})}\BibitemShut {NoStop}%
\bibitem [{\citenamefont {Aghababaei}\ \emph {et~al.}(2013)\citenamefont
  {Aghababaei}, \citenamefont {Haghighat},\ and\ \citenamefont
  {Kheirandish}}]{PRD.2013.87.047703}%
  \BibitemOpen
  \bibfield  {author} {\bibinfo {author} {\bibfnamefont {S.}~\bibnamefont
  {Aghababaei}}, \bibinfo {author} {\bibfnamefont {M.}~\bibnamefont
  {Haghighat}}, \ and\ \bibinfo {author} {\bibfnamefont {A.}~\bibnamefont
  {Kheirandish}},\ }\href {\doibase 10.1103/PhysRevD.87.047703} {\bibfield
  {journal} {\bibinfo  {journal} {Phys. Rev. D}\ }\textbf {\bibinfo {volume}
  {87}},\ \bibinfo {pages} {047703} (\bibinfo {year} {2013})}\BibitemShut
  {NoStop}%
\bibitem [{\citenamefont {Jackiw}\ and\ \citenamefont
  {Kosteleck\'{y}}(1999)}]{PRL.1999.82.3572}%
  \BibitemOpen
  \bibfield  {author} {\bibinfo {author} {\bibfnamefont {R.}~\bibnamefont
  {Jackiw}}\ and\ \bibinfo {author} {\bibfnamefont {V.~A.}\ \bibnamefont
  {Kosteleck\'{y}}},\ }\href {\doibase 10.1103/PhysRevLett.82.3572} {\bibfield
  {journal} {\bibinfo  {journal} {Phys. Rev. Lett.}\ }\textbf {\bibinfo
  {volume} {82}},\ \bibinfo {pages} {3572} (\bibinfo {year}
  {1999})}\BibitemShut {NoStop}%
\bibitem [{\citenamefont {P\'{e}rez-Victoria}(1999)}]{PRL.1999.83.2518}%
  \BibitemOpen
  \bibfield  {author} {\bibinfo {author} {\bibfnamefont {M.}~\bibnamefont
  {P\'{e}rez-Victoria}},\ }\href {\doibase 10.1103/PhysRevLett.83.2518}
  {\bibfield  {journal} {\bibinfo  {journal} {Phys. Rev. Lett.}\ }\textbf
  {\bibinfo {volume} {83}},\ \bibinfo {pages} {2518} (\bibinfo {year}
  {1999})}\BibitemShut {NoStop}%
\bibitem [{\citenamefont {Chung}\ and\ \citenamefont
  {Chung}(2001)}]{PRD.2001.63.105015}%
  \BibitemOpen
  \bibfield  {author} {\bibinfo {author} {\bibfnamefont {J.~M.}\ \bibnamefont
  {Chung}}\ and\ \bibinfo {author} {\bibfnamefont {B.~K.}\ \bibnamefont
  {Chung}},\ }\href {\doibase 10.1103/PhysRevD.63.105015} {\bibfield  {journal}
  {\bibinfo  {journal} {Phys. Rev. D}\ }\textbf {\bibinfo {volume} {63}},\
  \bibinfo {pages} {105015} (\bibinfo {year} {2001})}\BibitemShut {NoStop}%
\bibitem [{\citenamefont {Barreto}\ \emph {et~al.}(2006)\citenamefont
  {Barreto}, \citenamefont {Bazeia},\ and\ \citenamefont
  {Menezes}}]{PRD.2006.73.65015}%
  \BibitemOpen
  \bibfield  {author} {\bibinfo {author} {\bibfnamefont {M.~N.}\ \bibnamefont
  {Barreto}}, \bibinfo {author} {\bibfnamefont {D.}~\bibnamefont {Bazeia}}, \
  and\ \bibinfo {author} {\bibfnamefont {R.}~\bibnamefont {Menezes}},\ }\href
  {\doibase 10.1103/PhysRevD.73.065015} {\bibfield  {journal} {\bibinfo
  {journal} {Phys. Rev. D}\ }\textbf {\bibinfo {volume} {73}},\ \bibinfo
  {pages} {065015} (\bibinfo {year} {2006})}\BibitemShut {NoStop}%
\bibitem [{\citenamefont {Avelino}\ \emph {et~al.}(2009)\citenamefont
  {Avelino}, \citenamefont {Bazeia}, \citenamefont {Losano}, \citenamefont
  {Menezes},\ and\ \citenamefont {Rodrigues}}]{PRD.2009.79.123503}%
  \BibitemOpen
  \bibfield  {author} {\bibinfo {author} {\bibfnamefont {P.~P.}\ \bibnamefont
  {Avelino}}, \bibinfo {author} {\bibfnamefont {D.}~\bibnamefont {Bazeia}},
  \bibinfo {author} {\bibfnamefont {L.}~\bibnamefont {Losano}}, \bibinfo
  {author} {\bibfnamefont {R.}~\bibnamefont {Menezes}}, \ and\ \bibinfo
  {author} {\bibfnamefont {J.~J.}\ \bibnamefont {Rodrigues}},\ }\href {\doibase
  10.1103/PhysRevD.79.123503} {\bibfield  {journal} {\bibinfo  {journal} {Phys.
  Rev. D}\ }\textbf {\bibinfo {volume} {79}},\ \bibinfo {pages} {123503}
  (\bibinfo {year} {2009})}\BibitemShut {NoStop}%
\bibitem [{\citenamefont {Bazeia}\ \emph {et~al.}(2010)\citenamefont {Bazeia},
  \citenamefont {Ferreira~Jr.}, \citenamefont {Gomes},\ and\ \citenamefont
  {Menezes}}]{PD.2010.239.942}%
  \BibitemOpen
  \bibfield  {author} {\bibinfo {author} {\bibfnamefont {D.}~\bibnamefont
  {Bazeia}}, \bibinfo {author} {\bibfnamefont {M.}~\bibnamefont
  {Ferreira~Jr.}}, \bibinfo {author} {\bibfnamefont {A.}~\bibnamefont {Gomes}},
  \ and\ \bibinfo {author} {\bibfnamefont {R.}~\bibnamefont {Menezes}},\ }\href
  {\doibase 10.1016/j.physd.2010.01.015} {\bibfield  {journal} {\bibinfo
  {journal} {Phys. D}\ }\textbf {\bibinfo {volume} {239}},\ \bibinfo {pages}
  {942} (\bibinfo {year} {2010})}\BibitemShut {NoStop}%
\bibitem [{\citenamefont {Miller}\ \emph {et~al.}(2012)\citenamefont {Miller},
  \citenamefont {Casana}, \citenamefont {Ferreira~Jr.},\ and\ \citenamefont
  {da~Hora}}]{PRD.2012.86.065011}%
  \BibitemOpen
  \bibfield  {author} {\bibinfo {author} {\bibfnamefont {C.}~\bibnamefont
  {Miller}}, \bibinfo {author} {\bibfnamefont {R.}~\bibnamefont {Casana}},
  \bibinfo {author} {\bibfnamefont {M.~M.}\ \bibnamefont {Ferreira~Jr.}}, \
  and\ \bibinfo {author} {\bibfnamefont {E.}~\bibnamefont {da~Hora}},\ }\href
  {\doibase 10.1103/PhysRevD.86.065011} {\bibfield  {journal} {\bibinfo
  {journal} {Phys. Rev. D}\ }\textbf {\bibinfo {volume} {86}},\ \bibinfo
  {pages} {065011} (\bibinfo {year} {2012})}\BibitemShut {NoStop}%
\bibitem [{\citenamefont {Casana}\ \emph {et~al.}(2008)\citenamefont {Casana},
  \citenamefont {Ferreira~Jr.},\ and\ \citenamefont
  {Rodrigues}}]{PRD.2008.78.125013}%
  \BibitemOpen
  \bibfield  {author} {\bibinfo {author} {\bibfnamefont {R.}~\bibnamefont
  {Casana}}, \bibinfo {author} {\bibfnamefont {M.~M.}\ \bibnamefont
  {Ferreira~Jr.}}, \ and\ \bibinfo {author} {\bibfnamefont {J.~S.}\
  \bibnamefont {Rodrigues}},\ }\href {\doibase 10.1103/PhysRevD.78.125013}
  {\bibfield  {journal} {\bibinfo  {journal} {Phys. Rev. D}\ }\textbf {\bibinfo
  {volume} {78}},\ \bibinfo {pages} {125013} (\bibinfo {year}
  {2008})}\BibitemShut {NoStop}%
\bibitem [{\citenamefont {Altschul}(2011)}]{PRD.2011.84.076006}%
  \BibitemOpen
  \bibfield  {author} {\bibinfo {author} {\bibfnamefont {B.}~\bibnamefont
  {Altschul}},\ }\href {\doibase 10.1103/PhysRevD.84.076006} {\bibfield
  {journal} {\bibinfo  {journal} {Phys. Rev. D}\ }\textbf {\bibinfo {volume}
  {84}},\ \bibinfo {pages} {076006} (\bibinfo {year} {2011})}\BibitemShut
  {NoStop}%
\bibitem [{\citenamefont {Ganguly}\ \emph {et~al.}(2011)\citenamefont
  {Ganguly}, \citenamefont {Gangopadhyay},\ and\ \citenamefont
  {Majumdar}}]{EPL.2011.96.61001}%
  \BibitemOpen
  \bibfield  {author} {\bibinfo {author} {\bibfnamefont {O.}~\bibnamefont
  {Ganguly}}, \bibinfo {author} {\bibfnamefont {D.}~\bibnamefont
  {Gangopadhyay}}, \ and\ \bibinfo {author} {\bibfnamefont {P.}~\bibnamefont
  {Majumdar}},\ }\href@noop {} {\bibfield  {journal} {\bibinfo  {journal} {Eur.
  Phys. Lett.}\ }\textbf {\bibinfo {volume} {96}},\ \bibinfo {pages} {61001}
  (\bibinfo {year} {2011})}\BibitemShut {NoStop}%
\bibitem [{\citenamefont {Cambiaso}\ \emph {et~al.}(2012)\citenamefont
  {Cambiaso}, \citenamefont {Lehnert},\ and\ \citenamefont
  {Potting}}]{PRD.2012.85.085023}%
  \BibitemOpen
  \bibfield  {author} {\bibinfo {author} {\bibfnamefont {M.}~\bibnamefont
  {Cambiaso}}, \bibinfo {author} {\bibfnamefont {R.}~\bibnamefont {Lehnert}}, \
  and\ \bibinfo {author} {\bibfnamefont {R.}~\bibnamefont {Potting}},\ }\href
  {\doibase 10.1103/PhysRevD.85.085023} {\bibfield  {journal} {\bibinfo
  {journal} {Phys. Rev. D}\ }\textbf {\bibinfo {volume} {85}},\ \bibinfo
  {pages} {085023} (\bibinfo {year} {2012})}\BibitemShut {NoStop}%
\bibitem [{\citenamefont {Pospelov}\ and\ \citenamefont
  {Shang}(2012)}]{PRD.2012.85.105001}%
  \BibitemOpen
  \bibfield  {author} {\bibinfo {author} {\bibfnamefont {M.}~\bibnamefont
  {Pospelov}}\ and\ \bibinfo {author} {\bibfnamefont {Y.}~\bibnamefont
  {Shang}},\ }\href {\doibase 10.1103/PhysRevD.85.105001} {\bibfield  {journal}
  {\bibinfo  {journal} {Phys. Rev. D}\ }\textbf {\bibinfo {volume} {85}},\
  \bibinfo {pages} {105001} (\bibinfo {year} {2012})}\BibitemShut {NoStop}%
\bibitem [{\citenamefont {Casana}\ \emph {et~al.}(2011)\citenamefont {Casana},
  \citenamefont {Carvalho},\ and\ \citenamefont
  {Ferreira~Jr.}}]{PRD.2011.84.045008}%
  \BibitemOpen
  \bibfield  {author} {\bibinfo {author} {\bibfnamefont {R.}~\bibnamefont
  {Casana}}, \bibinfo {author} {\bibfnamefont {E.~S.}\ \bibnamefont
  {Carvalho}}, \ and\ \bibinfo {author} {\bibfnamefont {M.~M.}\ \bibnamefont
  {Ferreira~Jr.}},\ }\href {\doibase 10.1103/PhysRevD.84.045008} {\bibfield
  {journal} {\bibinfo  {journal} {Phys. Rev. D}\ }\textbf {\bibinfo {volume}
  {84}},\ \bibinfo {pages} {045008} (\bibinfo {year} {2011})}\BibitemShut
  {NoStop}%
\bibitem [{\citenamefont {Maluf}\ \emph {et~al.}(2014)\citenamefont {Maluf},
  \citenamefont {Almeida}, \citenamefont {Casana},\ and\ \citenamefont
  {Ferreira}}]{PRD.2014.90.025007}%
  \BibitemOpen
  \bibfield  {author} {\bibinfo {author} {\bibfnamefont {R.~V.}\ \bibnamefont
  {Maluf}}, \bibinfo {author} {\bibfnamefont {C.~A.~S.}\ \bibnamefont
  {Almeida}}, \bibinfo {author} {\bibfnamefont {R.}~\bibnamefont {Casana}}, \
  and\ \bibinfo {author} {\bibfnamefont {M.~M.}\ \bibnamefont {Ferreira}},\
  }\href {\doibase 10.1103/PhysRevD.90.025007} {\bibfield  {journal} {\bibinfo
  {journal} {Phys. Rev. D}\ }\textbf {\bibinfo {volume} {90}},\ \bibinfo
  {pages} {025007} (\bibinfo {year} {2014})}\BibitemShut {NoStop}%
\bibitem [{\citenamefont {Maluf}\ \emph {et~al.}(2013)\citenamefont {Maluf},
  \citenamefont {Santos}, \citenamefont {Cruz},\ and\ \citenamefont
  {Almeida}}]{PRD.2013.88.025005}%
  \BibitemOpen
  \bibfield  {author} {\bibinfo {author} {\bibfnamefont {R.~V.}\ \bibnamefont
  {Maluf}}, \bibinfo {author} {\bibfnamefont {V.}~\bibnamefont {Santos}},
  \bibinfo {author} {\bibfnamefont {W.~T.}\ \bibnamefont {Cruz}}, \ and\
  \bibinfo {author} {\bibfnamefont {C.~A.~S.}\ \bibnamefont {Almeida}},\ }\href
  {\doibase 10.1103/PhysRevD.88.025005} {\bibfield  {journal} {\bibinfo
  {journal} {Phys. Rev. D}\ }\textbf {\bibinfo {volume} {88}},\ \bibinfo
  {pages} {025005} (\bibinfo {year} {2013})}\BibitemShut {NoStop}%
\bibitem [{\citenamefont {Adam}\ and\ \citenamefont
  {Klinkhamer}(2003)}]{NPB.2003.657.214}%
  \BibitemOpen
  \bibfield  {author} {\bibinfo {author} {\bibfnamefont {C.}~\bibnamefont
  {Adam}}\ and\ \bibinfo {author} {\bibfnamefont {F.~R.}\ \bibnamefont
  {Klinkhamer}},\ }\href {\doibase 10.1016/S0550-3213(03)00143-3} {\bibfield
  {journal} {\bibinfo  {journal} {Nucl. Phys. B}\ }\textbf {\bibinfo {volume}
  {657}},\ \bibinfo {pages} {214} (\bibinfo {year} {2003})}\BibitemShut
  {NoStop}%
\bibitem [{\citenamefont {Andrianov}\ \emph {et~al.}(1998)\citenamefont
  {Andrianov}, \citenamefont {Soldati},\ and\ \citenamefont
  {Sorbo}}]{PRD.1998.59.25002}%
  \BibitemOpen
  \bibfield  {author} {\bibinfo {author} {\bibfnamefont {A.~A.}\ \bibnamefont
  {Andrianov}}, \bibinfo {author} {\bibfnamefont {R.}~\bibnamefont {Soldati}},
  \ and\ \bibinfo {author} {\bibfnamefont {L.}~\bibnamefont {Sorbo}},\ }\href
  {\doibase 10.1103/PhysRevD.59.025002} {\bibfield  {journal} {\bibinfo
  {journal} {Phys. Rev. D}\ }\textbf {\bibinfo {volume} {59}},\ \bibinfo
  {pages} {025002} (\bibinfo {year} {1998})}\BibitemShut {NoStop}%
\bibitem [{\citenamefont {Belich}\ \emph {et~al.}(2003)\citenamefont {Belich},
  \citenamefont {Ferreira~Jr.}, \citenamefont {Helay\"el-Neto},\ and\
  \citenamefont {Orlando}}]{PRD.2003.67.125011}%
  \BibitemOpen
  \bibfield  {author} {\bibinfo {author} {\bibfnamefont {H.}~\bibnamefont
  {Belich}}, \bibinfo {author} {\bibfnamefont {M.~M.}\ \bibnamefont
  {Ferreira~Jr.}}, \bibinfo {author} {\bibfnamefont {J.~A.}\ \bibnamefont
  {Helay\"el-Neto}}, \ and\ \bibinfo {author} {\bibfnamefont {M.~T.~D.}\
  \bibnamefont {Orlando}},\ }\href {\doibase 10.1103/PhysRevD.67.125011}
  {\bibfield  {journal} {\bibinfo  {journal} {Phys. Rev. D}\ }\textbf {\bibinfo
  {volume} {67}},\ \bibinfo {pages} {125011} (\bibinfo {year}
  {2003})}\BibitemShut {NoStop}%
\bibitem [{\citenamefont {Casana}\ \emph {et~al.}(2009)\citenamefont {Casana},
  \citenamefont {Ferreira~Jr.}, \citenamefont {Gomes},\ and\ \citenamefont
  {Pinheiro}}]{PRD.2009.80.125040}%
  \BibitemOpen
  \bibfield  {author} {\bibinfo {author} {\bibfnamefont {R.}~\bibnamefont
  {Casana}}, \bibinfo {author} {\bibfnamefont {M.~M.}\ \bibnamefont
  {Ferreira~Jr.}}, \bibinfo {author} {\bibfnamefont {A.~R.}\ \bibnamefont
  {Gomes}}, \ and\ \bibinfo {author} {\bibfnamefont {P.~R.~D.}\ \bibnamefont
  {Pinheiro}},\ }\href {\doibase 10.1103/PhysRevD.80.125040} {\bibfield
  {journal} {\bibinfo  {journal} {Phys. Rev. D}\ }\textbf {\bibinfo {volume}
  {80}},\ \bibinfo {pages} {125040} (\bibinfo {year} {2009})}\BibitemShut
  {NoStop}%
\bibitem [{\citenamefont {Lan}\ and\ \citenamefont
  {Wu}(2014)}]{EPJC.2014.74.2875}%
  \BibitemOpen
  \bibfield  {author} {\bibinfo {author} {\bibfnamefont {S.-q.}\ \bibnamefont
  {Lan}}\ and\ \bibinfo {author} {\bibfnamefont {F.}~\bibnamefont {Wu}},\
  }\href {\doibase 10.1140/epjc/s10052-014-2875-6} {\bibfield  {journal}
  {\bibinfo  {journal} {Eur. Phys. J. C}\ }\textbf {\bibinfo {volume} {74}},\
  \bibinfo {pages} {2875} (\bibinfo {year} {2014})}\BibitemShut {NoStop}%
\bibitem [{\citenamefont {Bufalo}(2014)}]{IJMPA.2014.29.1450112}%
  \BibitemOpen
  \bibfield  {author} {\bibinfo {author} {\bibfnamefont {R.}~\bibnamefont
  {Bufalo}},\ }\href {\doibase 10.1142/S0217751X14501127} {\bibfield  {journal}
  {\bibinfo  {journal} {Int. J. Mod. Phys. A}\ }\textbf {\bibinfo {volume}
  {29}},\ \bibinfo {pages} {1450112} (\bibinfo {year} {2014})}\BibitemShut
  {NoStop}%
\bibitem [{\citenamefont {Charneski}\ \emph {et~al.}(2012)\citenamefont
  {Charneski}, \citenamefont {Gomes}, \citenamefont {Maluf},\ and\
  \citenamefont {da~Silva}}]{PRD.2012.86.045003}%
  \BibitemOpen
  \bibfield  {author} {\bibinfo {author} {\bibfnamefont {B.}~\bibnamefont
  {Charneski}}, \bibinfo {author} {\bibfnamefont {M.}~\bibnamefont {Gomes}},
  \bibinfo {author} {\bibfnamefont {R.~V.}\ \bibnamefont {Maluf}}, \ and\
  \bibinfo {author} {\bibfnamefont {A.~J.}\ \bibnamefont {da~Silva}},\ }\href
  {\doibase 10.1103/PhysRevD.86.045003} {\bibfield  {journal} {\bibinfo
  {journal} {Phys. Rev. D}\ }\textbf {\bibinfo {volume} {86}},\ \bibinfo
  {pages} {045003} (\bibinfo {year} {2012})}\BibitemShut {NoStop}%
\bibitem [{\citenamefont {Casana}\ \emph {et~al.}(2010)\citenamefont {Casana},
  \citenamefont {Ferreira~Jr.},\ and\ \citenamefont
  {Silva}}]{PRD.2010.81.105015}%
  \BibitemOpen
  \bibfield  {author} {\bibinfo {author} {\bibfnamefont {R.}~\bibnamefont
  {Casana}}, \bibinfo {author} {\bibfnamefont {M.~M.}\ \bibnamefont
  {Ferreira~Jr.}}, \ and\ \bibinfo {author} {\bibfnamefont {M.~R.~O.}\
  \bibnamefont {Silva}},\ }\href {\doibase 10.1103/PhysRevD.81.105015}
  {\bibfield  {journal} {\bibinfo  {journal} {Phys. Rev. D}\ }\textbf {\bibinfo
  {volume} {81}},\ \bibinfo {pages} {105015} (\bibinfo {year}
  {2010})}\BibitemShut {NoStop}%
\bibitem [{\citenamefont {Casana}\ \emph {et~al.}(2012)\citenamefont {Casana},
  \citenamefont {Ferreira~Jr.},\ and\ \citenamefont
  {Moreira}}]{EPJC.2012.72.2070}%
  \BibitemOpen
  \bibfield  {author} {\bibinfo {author} {\bibfnamefont {R.}~\bibnamefont
  {Casana}}, \bibinfo {author} {\bibfnamefont {M.~M.}\ \bibnamefont
  {Ferreira~Jr.}}, \ and\ \bibinfo {author} {\bibfnamefont {R.~P.~M.}\
  \bibnamefont {Moreira}},\ }\href {\doibase 10.1140/epjc/s10052-012-2070-6}
  {\bibfield  {journal} {\bibinfo  {journal} {Eur. Phys. J. C}\ }\textbf
  {\bibinfo {volume} {72}},\ \bibinfo {pages} {2070} (\bibinfo {year}
  {2012})}\BibitemShut {NoStop}%
\bibitem [{\citenamefont {Belich}\ and\ \citenamefont
  {Bakke}(2014)}]{PRD.2014.90.025026}%
  \BibitemOpen
  \bibfield  {author} {\bibinfo {author} {\bibfnamefont {H.}~\bibnamefont
  {Belich}}\ and\ \bibinfo {author} {\bibfnamefont {K.}~\bibnamefont {Bakke}},\
  }\href {\doibase 10.1103/PhysRevD.90.025026} {\bibfield  {journal} {\bibinfo
  {journal} {Phys. Rev. D}\ }\textbf {\bibinfo {volume} {90}},\ \bibinfo
  {pages} {025026} (\bibinfo {year} {2014})}\BibitemShut {NoStop}%
\bibitem [{\citenamefont {Bakke}\ and\ \citenamefont
  {Belich}(2013)}]{AoP.2013.333.272}%
  \BibitemOpen
  \bibfield  {author} {\bibinfo {author} {\bibfnamefont {K.}~\bibnamefont
  {Bakke}}\ and\ \bibinfo {author} {\bibfnamefont {H.}~\bibnamefont {Belich}},\
  }\href {\doibase http://dx.doi.org/10.1016/j.aop.2013.03.009} {\bibfield
  {journal} {\bibinfo  {journal} {Ann. Phys. (N.Y.)}\ }\textbf {\bibinfo
  {volume} {333}},\ \bibinfo {pages} {272 } (\bibinfo {year}
  {2013})}\BibitemShut {NoStop}%
\bibitem [{\citenamefont {Silva}\ and\ \citenamefont
  {Andrade}(2013)}]{EPL.2013.101.51005}%
  \BibitemOpen
  \bibfield  {author} {\bibinfo {author} {\bibfnamefont {E.~O.}\ \bibnamefont
  {Silva}}\ and\ \bibinfo {author} {\bibfnamefont {F.~M.}\ \bibnamefont
  {Andrade}},\ }\href {\doibase 10.1209/0295-5075/101/51005} {\bibfield
  {journal} {\bibinfo  {journal} {Europhys. Lett.}\ }\textbf {\bibinfo {volume}
  {101}},\ \bibinfo {pages} {51005} (\bibinfo {year} {2013})}\BibitemShut
  {NoStop}%
\bibitem [{\citenamefont {Bakke}\ \emph
  {et~al.}(2011{\natexlab{a}})\citenamefont {Bakke}, \citenamefont {Belich},\
  and\ \citenamefont {O.~Silva}}]{ADP.2011.523.910}%
  \BibitemOpen
  \bibfield  {author} {\bibinfo {author} {\bibfnamefont {K.}~\bibnamefont
  {Bakke}}, \bibinfo {author} {\bibfnamefont {H.}~\bibnamefont {Belich}}, \
  and\ \bibinfo {author} {\bibfnamefont {E.}~\bibnamefont {O.~Silva}},\ }\href
  {\doibase 10.1002/andp.201100087} {\bibfield  {journal} {\bibinfo  {journal}
  {Ann. Phys. (Berlin)}\ }\textbf {\bibinfo {volume} {523}},\ \bibinfo {pages}
  {910} (\bibinfo {year} {2011}{\natexlab{a}})}\BibitemShut {NoStop}%
\bibitem [{\citenamefont {Bakke}\ \emph {et~al.}(2012)\citenamefont {Bakke},
  \citenamefont {Silva},\ and\ \citenamefont {Belich}}]{JPG.2012.39.55004}%
  \BibitemOpen
  \bibfield  {author} {\bibinfo {author} {\bibfnamefont {K.}~\bibnamefont
  {Bakke}}, \bibinfo {author} {\bibfnamefont {E.~O.}\ \bibnamefont {Silva}}, \
  and\ \bibinfo {author} {\bibfnamefont {H.}~\bibnamefont {Belich}},\ }\href
  {\doibase 10.1088/0954-3899/39/5/055004} {\bibfield  {journal} {\bibinfo
  {journal} {J. Phys. G: Nucl. Part. Phys.}\ }\textbf {\bibinfo {volume}
  {39}},\ \bibinfo {pages} {055004} (\bibinfo {year} {2012})}\BibitemShut
  {NoStop}%
\bibitem [{\citenamefont {Bakke}\ \emph
  {et~al.}(2011{\natexlab{b}})\citenamefont {Bakke}, \citenamefont {Belich},\
  and\ \citenamefont {Silva}}]{JMP.2011.52.063505}%
  \BibitemOpen
  \bibfield  {author} {\bibinfo {author} {\bibfnamefont {K.}~\bibnamefont
  {Bakke}}, \bibinfo {author} {\bibfnamefont {H.}~\bibnamefont {Belich}}, \
  and\ \bibinfo {author} {\bibfnamefont {E.~O.}\ \bibnamefont {Silva}},\ }\href
  {\doibase 10.1063/1.3597230} {\bibfield  {journal} {\bibinfo  {journal} {J.
  Math. Phys.}\ }\textbf {\bibinfo {volume} {52}},\ \bibinfo {pages} {063505}
  (\bibinfo {year} {2011}{\natexlab{b}})}\BibitemShut {NoStop}%
\bibitem [{\citenamefont {Belich}\ \emph {et~al.}(2011)\citenamefont {Belich},
  \citenamefont {Silva}, \citenamefont {Ferreira~Jr.},\ and\ \citenamefont
  {Orlando}}]{PRD.2011.83.125025}%
  \BibitemOpen
  \bibfield  {author} {\bibinfo {author} {\bibfnamefont {H.}~\bibnamefont
  {Belich}}, \bibinfo {author} {\bibfnamefont {E.~O.}\ \bibnamefont {Silva}},
  \bibinfo {author} {\bibfnamefont {M.~M.}\ \bibnamefont {Ferreira~Jr.}}, \
  and\ \bibinfo {author} {\bibfnamefont {M.~T.~D.}\ \bibnamefont {Orlando}},\
  }\href {\doibase 10.1103/PhysRevD.83.125025} {\bibfield  {journal} {\bibinfo
  {journal} {Phys. Rev. D}\ }\textbf {\bibinfo {volume} {83}},\ \bibinfo
  {pages} {125025} (\bibinfo {year} {2011})}\BibitemShut {NoStop}%
\bibitem [{\citenamefont {Jacobson}\ \emph {et~al.}(2002)\citenamefont
  {Jacobson}, \citenamefont {Liberati},\ and\ \citenamefont
  {Mattingly}}]{PRD.2002.66.081302}%
  \BibitemOpen
  \bibfield  {author} {\bibinfo {author} {\bibfnamefont {T.}~\bibnamefont
  {Jacobson}}, \bibinfo {author} {\bibfnamefont {S.}~\bibnamefont {Liberati}},
  \ and\ \bibinfo {author} {\bibfnamefont {D.}~\bibnamefont {Mattingly}},\
  }\href {\doibase 10.1103/PhysRevD.66.081302} {\bibfield  {journal} {\bibinfo
  {journal} {Phys. Rev. D}\ }\textbf {\bibinfo {volume} {66}},\ \bibinfo
  {pages} {081302} (\bibinfo {year} {2002})}\BibitemShut {NoStop}%
\bibitem [{\citenamefont {Liberati}\ and\ \citenamefont
  {Maccione}(2009)}]{AR.2009.59.245}%
  \BibitemOpen
  \bibfield  {author} {\bibinfo {author} {\bibfnamefont {S.}~\bibnamefont
  {Liberati}}\ and\ \bibinfo {author} {\bibfnamefont {L.}~\bibnamefont
  {Maccione}},\ }\href {\doibase 10.1146/annurev.nucl.010909.083640} {\bibfield
   {journal} {\bibinfo  {journal} {Ann. Rev. Nuclear Particle Science}\
  }\textbf {\bibinfo {volume} {59}},\ \bibinfo {pages} {245} (\bibinfo {year}
  {2009})}\BibitemShut {NoStop}%
\bibitem [{\citenamefont {Shao}\ and\ \citenamefont
  {Ma}(2011)}]{PRD.2011.83.127702}%
  \BibitemOpen
  \bibfield  {author} {\bibinfo {author} {\bibfnamefont {L.}~\bibnamefont
  {Shao}}\ and\ \bibinfo {author} {\bibfnamefont {B.}~\bibnamefont {Ma}},\
  }\href {\doibase 10.1103/PhysRevD.83.127702} {\bibfield  {journal} {\bibinfo
  {journal} {Phys. Rev. D}\ }\textbf {\bibinfo {volume} {83}},\ \bibinfo
  {pages} {127702} (\bibinfo {year} {2011})}\BibitemShut {NoStop}%
\bibitem [{\citenamefont {Kosteleck\'{y}}\ and\ \citenamefont
  {Russell}(2011)}]{RMP.2011.83.11}%
  \BibitemOpen
  \bibfield  {author} {\bibinfo {author} {\bibfnamefont {V.~A.}\ \bibnamefont
  {Kosteleck\'{y}}}\ and\ \bibinfo {author} {\bibfnamefont {N.}~\bibnamefont
  {Russell}},\ }\href {\doibase 10.1103/RevModPhys.83.11} {\bibfield  {journal}
  {\bibinfo  {journal} {Rev. Mod. Phys.}\ }\textbf {\bibinfo {volume} {83}},\
  \bibinfo {pages} {11} (\bibinfo {year} {2011})}\BibitemShut {NoStop}%
\bibitem [{\citenamefont {Bluhm}\ \emph {et~al.}(2002)\citenamefont {Bluhm},
  \citenamefont {Kosteleck\'y}, \citenamefont {Lane},\ and\ \citenamefont
  {Russell}}]{PRL.2002.88.090801}%
  \BibitemOpen
  \bibfield  {author} {\bibinfo {author} {\bibfnamefont {R.}~\bibnamefont
  {Bluhm}}, \bibinfo {author} {\bibfnamefont {V.~A.}\ \bibnamefont
  {Kosteleck\'y}}, \bibinfo {author} {\bibfnamefont {C.~D.}\ \bibnamefont
  {Lane}}, \ and\ \bibinfo {author} {\bibfnamefont {N.}~\bibnamefont
  {Russell}},\ }\href {\doibase 10.1103/PhysRevLett.88.090801} {\bibfield
  {journal} {\bibinfo  {journal} {Phys. Rev. Lett.}\ }\textbf {\bibinfo
  {volume} {88}},\ \bibinfo {pages} {090801} (\bibinfo {year}
  {2002})}\BibitemShut {NoStop}%
\bibitem [{\citenamefont {Bluhm}\ and\ \citenamefont
  {Kosteleck\'y}(2000)}]{PRL.2000.84.1381}%
  \BibitemOpen
  \bibfield  {author} {\bibinfo {author} {\bibfnamefont {R.}~\bibnamefont
  {Bluhm}}\ and\ \bibinfo {author} {\bibfnamefont {V.~A.}\ \bibnamefont
  {Kosteleck\'y}},\ }\href {\doibase 10.1103/PhysRevLett.84.1381} {\bibfield
  {journal} {\bibinfo  {journal} {Phys. Rev. Lett.}\ }\textbf {\bibinfo
  {volume} {84}},\ \bibinfo {pages} {1381} (\bibinfo {year}
  {2000})}\BibitemShut {NoStop}%
\bibitem [{\citenamefont {Belich}\ \emph {et~al.}(2005)\citenamefont {Belich},
  \citenamefont {Costa-Soares}, \citenamefont {Ferreira~Jr.},\ and\
  \citenamefont {Helayël-Neto}}]{EPJC.2005.41.421}%
  \BibitemOpen
  \bibfield  {author} {\bibinfo {author} {\bibfnamefont {H.}~\bibnamefont
  {Belich}}, \bibinfo {author} {\bibfnamefont {T.}~\bibnamefont
  {Costa-Soares}}, \bibinfo {author} {\bibfnamefont {M.~M.}\ \bibnamefont
  {Ferreira~Jr.}}, \ and\ \bibinfo {author} {\bibfnamefont {J.~A.}\
  \bibnamefont {Helayël-Neto}},\ }\href {\doibase 10.1140/epjc/s2005-02240-y}
  {\bibfield  {journal} {\bibinfo  {journal} {Eur. Phys. J. C}\ }\textbf
  {\bibinfo {volume} {41}},\ \bibinfo {pages} {421} (\bibinfo {year}
  {2005})}\BibitemShut {NoStop}%
\bibitem [{\citenamefont {Casana}\ \emph {et~al.}(2013)\citenamefont {Casana},
  \citenamefont {Ferreira~Jr.}, \citenamefont {Passos}, \citenamefont {dos
  Santos},\ and\ \citenamefont {Silva}}]{PRD.2013.87.047701}%
  \BibitemOpen
  \bibfield  {author} {\bibinfo {author} {\bibfnamefont {R.}~\bibnamefont
  {Casana}}, \bibinfo {author} {\bibfnamefont {M.~M.}\ \bibnamefont
  {Ferreira~Jr.}}, \bibinfo {author} {\bibfnamefont {E.}~\bibnamefont
  {Passos}}, \bibinfo {author} {\bibfnamefont {F.~E.~P.}\ \bibnamefont {dos
  Santos}}, \ and\ \bibinfo {author} {\bibfnamefont {E.~O.}\ \bibnamefont
  {Silva}},\ }\href {\doibase 10.1103/PhysRevD.87.047701} {\bibfield  {journal}
  {\bibinfo  {journal} {Phys. Rev. D}\ }\textbf {\bibinfo {volume} {87}},\
  \bibinfo {pages} {047701} (\bibinfo {year} {2013})}\BibitemShut {NoStop}%
\bibitem [{\citenamefont {Andrade}\ \emph
  {et~al.}(2013{\natexlab{a}})\citenamefont {Andrade}, \citenamefont {Silva},
  \citenamefont {Prud\^{e}ncio},\ and\ \citenamefont
  {Filgueiras}}]{JPG.2013.40.075007}%
  \BibitemOpen
  \bibfield  {author} {\bibinfo {author} {\bibfnamefont {F.~M.}\ \bibnamefont
  {Andrade}}, \bibinfo {author} {\bibfnamefont {E.~O.}\ \bibnamefont {Silva}},
  \bibinfo {author} {\bibfnamefont {T.}~\bibnamefont {Prud\^{e}ncio}}, \ and\
  \bibinfo {author} {\bibfnamefont {C.}~\bibnamefont {Filgueiras}},\ }\href
  {\doibase 10.1088/0954-3899/40/7/075007} {\bibfield  {journal} {\bibinfo
  {journal} {J. Phys. G}\ }\textbf {\bibinfo {volume} {40}},\ \bibinfo {pages}
  {075007} (\bibinfo {year} {2013}{\natexlab{a}})}\BibitemShut {NoStop}%
\bibitem [{\citenamefont {Andrade}\ \emph {et~al.}(2012)\citenamefont
  {Andrade}, \citenamefont {Silva},\ and\ \citenamefont
  {Pereira}}]{PRD.2012.85.041701}%
  \BibitemOpen
  \bibfield  {author} {\bibinfo {author} {\bibfnamefont {F.~M.}\ \bibnamefont
  {Andrade}}, \bibinfo {author} {\bibfnamefont {E.~O.}\ \bibnamefont {Silva}},
  \ and\ \bibinfo {author} {\bibfnamefont {M.}~\bibnamefont {Pereira}},\ }\href
  {\doibase 10.1103/PhysRevD.85.041701} {\bibfield  {journal} {\bibinfo
  {journal} {Phys. Rev. D}\ }\textbf {\bibinfo {volume} {85}},\ \bibinfo
  {pages} {041701(R)} (\bibinfo {year} {2012})}\BibitemShut {NoStop}%
\bibitem [{\citenamefont {Andrade}\ \emph
  {et~al.}(2013{\natexlab{b}})\citenamefont {Andrade}, \citenamefont {Silva},\
  and\ \citenamefont {Pereira}}]{AoP.2013.339.510}%
  \BibitemOpen
  \bibfield  {author} {\bibinfo {author} {\bibfnamefont {F.~M.}\ \bibnamefont
  {Andrade}}, \bibinfo {author} {\bibfnamefont {E.~O.}\ \bibnamefont {Silva}},
  \ and\ \bibinfo {author} {\bibfnamefont {M.}~\bibnamefont {Pereira}},\ }\href
  {\doibase 10.1016/j.aop.2013.10.001} {\bibfield  {journal} {\bibinfo
  {journal} {Ann. Phys. (N.Y.)}\ }\textbf {\bibinfo {volume} {339}},\ \bibinfo
  {pages} {510} (\bibinfo {year} {2013}{\natexlab{b}})}\BibitemShut {NoStop}%
\bibitem [{\citenamefont {Bulla}\ and\ \citenamefont
  {Gesztesy}(1985)}]{JMP.1985.26.2520}%
  \BibitemOpen
  \bibfield  {author} {\bibinfo {author} {\bibfnamefont {W.}~\bibnamefont
  {Bulla}}\ and\ \bibinfo {author} {\bibfnamefont {F.}~\bibnamefont
  {Gesztesy}},\ }\href {\doibase 10.1063/1.526768} {\bibfield  {journal}
  {\bibinfo  {journal} {J. Math. Phys.}\ }\textbf {\bibinfo {volume} {26}},\
  \bibinfo {pages} {2520} (\bibinfo {year} {1985})}\BibitemShut {NoStop}%
\bibitem [{\citenamefont {Kay}\ and\ \citenamefont
  {Studer}(1991)}]{CMP.1991.139.103}%
  \BibitemOpen
  \bibfield  {author} {\bibinfo {author} {\bibfnamefont {B.~S.}\ \bibnamefont
  {Kay}}\ and\ \bibinfo {author} {\bibfnamefont {U.~M.}\ \bibnamefont
  {Studer}},\ }\href {\doibase 10.1007/BF02102731} {\bibfield  {journal}
  {\bibinfo  {journal} {Commun. Math. Phys.}\ }\textbf {\bibinfo {volume}
  {139}},\ \bibinfo {pages} {103} (\bibinfo {year} {1991})}\BibitemShut
  {NoStop}%
\bibitem [{\citenamefont {Albeverio}\ \emph {et~al.}(2004)\citenamefont
  {Albeverio}, \citenamefont {Gesztesy}, \citenamefont {Hoegh-Krohn},\ and\
  \citenamefont {Holden}}]{Book.2004.Albeverio}%
  \BibitemOpen
  \bibfield  {author} {\bibinfo {author} {\bibfnamefont {S.}~\bibnamefont
  {Albeverio}}, \bibinfo {author} {\bibfnamefont {F.}~\bibnamefont {Gesztesy}},
  \bibinfo {author} {\bibfnamefont {R.}~\bibnamefont {Hoegh-Krohn}}, \ and\
  \bibinfo {author} {\bibfnamefont {H.}~\bibnamefont {Holden}},\ }\href
  {http://books.google.com.br/books?id=cR6uhyqnpUQC} {\emph {\bibinfo {title}
  {Solvable Models in Quantum Mechanics}}},\ \bibinfo {edition} {2nd}\ ed.\
  (\bibinfo  {publisher} {AMS Chelsea Publishing},\ \bibinfo {address}
  {Providence, RI},\ \bibinfo {year} {2004})\BibitemShut {NoStop}%
\bibitem [{\citenamefont {Hagen}(1990{\natexlab{a}})}]{PRL.1990.64.2347}%
  \BibitemOpen
  \bibfield  {author} {\bibinfo {author} {\bibfnamefont {C.~R.}\ \bibnamefont
  {Hagen}},\ }\href {\doibase 10.1103/PhysRevLett.64.2347} {\bibfield
  {journal} {\bibinfo  {journal} {Phys. Rev. Lett.}\ }\textbf {\bibinfo
  {volume} {64}},\ \bibinfo {pages} {2347} (\bibinfo {year}
  {1990}{\natexlab{a}})}\BibitemShut {NoStop}%
\bibitem [{\citenamefont {Hagen}(1990{\natexlab{b}})}]{PRL.1990.64.503}%
  \BibitemOpen
  \bibfield  {author} {\bibinfo {author} {\bibfnamefont {C.~R.}\ \bibnamefont
  {Hagen}},\ }\href {\doibase 10.1103/PhysRevLett.64.503} {\bibfield  {journal}
  {\bibinfo  {journal} {Phys. Rev. Lett.}\ }\textbf {\bibinfo {volume} {64}},\
  \bibinfo {pages} {503} (\bibinfo {year} {1990}{\natexlab{b}})}\BibitemShut
  {NoStop}%
\bibitem [{\citenamefont {Ruijsenaars}(1983)}]{AoP.1983.146.1}%
  \BibitemOpen
  \bibfield  {author} {\bibinfo {author} {\bibfnamefont {S.~N.~M.}\
  \bibnamefont {Ruijsenaars}},\ }\href {\doibase 10.1016/0003-4916(83)90051-9}
  {\bibfield  {journal} {\bibinfo  {journal} {Ann. Phys. (NY)}\ }\textbf
  {\bibinfo {volume} {146}},\ \bibinfo {pages} {1} (\bibinfo {year}
  {1983})}\BibitemShut {NoStop}%
\bibitem [{\citenamefont {Adami}\ and\ \citenamefont
  {Teta}(1998)}]{LMP.1998.43.43}%
  \BibitemOpen
  \bibfield  {author} {\bibinfo {author} {\bibfnamefont {R.}~\bibnamefont
  {Adami}}\ and\ \bibinfo {author} {\bibfnamefont {A.}~\bibnamefont {Teta}},\
  }\href {\doibase 10.1023/A:1007330512611} {\bibfield  {journal} {\bibinfo
  {journal} {Lett. Math. Phys.}\ }\textbf {\bibinfo {volume} {43}},\ \bibinfo
  {pages} {43} (\bibinfo {year} {1998})}\BibitemShut {NoStop}%
\bibitem [{\citenamefont {Dabrowski}\ and\ \citenamefont
  {Stovicek}(1998)}]{JMP.1998.39.47}%
  \BibitemOpen
  \bibfield  {author} {\bibinfo {author} {\bibfnamefont {L.}~\bibnamefont
  {Dabrowski}}\ and\ \bibinfo {author} {\bibfnamefont {P.}~\bibnamefont
  {Stovicek}},\ }\href {\doibase 10.1063/1.532307} {\bibfield  {journal}
  {\bibinfo  {journal} {J. Math. Phys.}\ }\textbf {\bibinfo {volume} {39}},\
  \bibinfo {pages} {47} (\bibinfo {year} {1998})}\BibitemShut {NoStop}%
\bibitem [{\citenamefont {Reed}\ and\ \citenamefont
  {Simon}(1975)}]{Book.1975.Reed.II}%
  \BibitemOpen
  \bibfield  {author} {\bibinfo {author} {\bibfnamefont {M.}~\bibnamefont
  {Reed}}\ and\ \bibinfo {author} {\bibfnamefont {B.}~\bibnamefont {Simon}},\
  }\href@noop {} {\emph {\bibinfo {title} {Methods of Modern Mathematical
  Physics. II. Fourier Analysis, Self-Adjointness.}}}\ (\bibinfo  {publisher}
  {Academic Press},\ \bibinfo {address} {New York - London},\ \bibinfo {year}
  {1975})\BibitemShut {NoStop}%
\bibitem [{\citenamefont {Bordag}\ and\ \citenamefont
  {Voropaev}(1994)}]{PLB.1994.333.238}%
  \BibitemOpen
  \bibfield  {author} {\bibinfo {author} {\bibfnamefont {M.}~\bibnamefont
  {Bordag}}\ and\ \bibinfo {author} {\bibfnamefont {S.}~\bibnamefont
  {Voropaev}},\ }\href {\doibase 10.1016/0370-2693(94)91037-5} {\bibfield
  {journal} {\bibinfo  {journal} {Phys. Lett. B}\ }\textbf {\bibinfo {volume}
  {333}},\ \bibinfo {pages} {238} (\bibinfo {year} {1994})}\BibitemShut
  {NoStop}%
\bibitem [{\citenamefont {Sakurai}\ and\ \citenamefont
  {Napolitano}(2011)}]{Book.2011.Sakurai}%
  \BibitemOpen
  \bibfield  {author} {\bibinfo {author} {\bibfnamefont {J.~J.}\ \bibnamefont
  {Sakurai}}\ and\ \bibinfo {author} {\bibfnamefont {J.}~\bibnamefont
  {Napolitano}},\ }\href@noop {} {\emph {\bibinfo {title} {Modern Quantum
  Mechanics}}},\ \bibinfo {edition} {2nd}\ ed.\ (\bibinfo  {publisher}
  {Addison-Wesley},\ \bibinfo {year} {2011})\BibitemShut {NoStop}%
\bibitem [{\citenamefont {Audretsch}\ \emph {et~al.}(1995)\citenamefont
  {Audretsch}, \citenamefont {Jasper},\ and\ \citenamefont
  {Skarzhinsky}}]{JPA.1995.28.2359}%
  \BibitemOpen
  \bibfield  {author} {\bibinfo {author} {\bibfnamefont {J.}~\bibnamefont
  {Audretsch}}, \bibinfo {author} {\bibfnamefont {U.}~\bibnamefont {Jasper}}, \
  and\ \bibinfo {author} {\bibfnamefont {V.~D.}\ \bibnamefont {Skarzhinsky}},\
  }\href {\doibase 10.1088/0305-4470/28/8/026} {\bibfield  {journal} {\bibinfo
  {journal} {J. Phys. A}\ }\textbf {\bibinfo {volume} {28}},\ \bibinfo {pages}
  {2359} (\bibinfo {year} {1995})}\BibitemShut {NoStop}%
\bibitem [{\citenamefont {Coutinho}\ \emph {et~al.}(1992)\citenamefont
  {Coutinho}, \citenamefont {Nogami},\ and\ \citenamefont
  {Fernando~Perez}}]{PRA.1992.46.6052}%
  \BibitemOpen
  \bibfield  {author} {\bibinfo {author} {\bibfnamefont {F.~A.~B.}\
  \bibnamefont {Coutinho}}, \bibinfo {author} {\bibfnamefont {Y.}~\bibnamefont
  {Nogami}}, \ and\ \bibinfo {author} {\bibfnamefont {J.}~\bibnamefont
  {Fernando~Perez}},\ }\href {\doibase 10.1103/PhysRevA.46.6052} {\bibfield
  {journal} {\bibinfo  {journal} {Phys. Rev. A}\ }\textbf {\bibinfo {volume}
  {46}},\ \bibinfo {pages} {6052} (\bibinfo {year} {1992})}\BibitemShut
  {NoStop}%
\bibitem [{\citenamefont {de~Oliveira}\ and\ \citenamefont
  {Pereira}(2010)}]{JPA.2010.43.354011}%
  \BibitemOpen
  \bibfield  {author} {\bibinfo {author} {\bibfnamefont {C.~R.}\ \bibnamefont
  {de~Oliveira}}\ and\ \bibinfo {author} {\bibfnamefont {M.}~\bibnamefont
  {Pereira}},\ }\href {\doibase 10.1088/1751-8113/43/35/354011} {\bibfield
  {journal} {\bibinfo  {journal} {J. Phys. A}\ }\textbf {\bibinfo {volume}
  {43}},\ \bibinfo {pages} {354011} (\bibinfo {year} {2010})}\BibitemShut
  {NoStop}%
\bibitem [{\citenamefont {Abramowitz}\ and\ \citenamefont
  {Stegun}(1972)}]{Book.1972.Abramowitz}%
  \BibitemOpen
  \bibinfo {editor} {\bibfnamefont {M.}~\bibnamefont {Abramowitz}}\ and\
  \bibinfo {editor} {\bibfnamefont {I.~A.}\ \bibnamefont {Stegun}},\ eds.,\
  \href@noop {} {\emph {\bibinfo {title} {Handbook of Mathematical
  Functions}}}\ (\bibinfo  {publisher} {New York: Dover Publications},\
  \bibinfo {year} {1972})\BibitemShut {NoStop}%
\bibitem [{\citenamefont {Bennaceur}\ \emph {et~al.}(1999)\citenamefont
  {Bennaceur}, \citenamefont {Dobaczewski},\ and\ \citenamefont
  {Ploszajczak}}]{PRC.1999.60.34308}%
  \BibitemOpen
  \bibfield  {author} {\bibinfo {author} {\bibfnamefont {K.}~\bibnamefont
  {Bennaceur}}, \bibinfo {author} {\bibfnamefont {J.}~\bibnamefont
  {Dobaczewski}}, \ and\ \bibinfo {author} {\bibfnamefont {M.}~\bibnamefont
  {Ploszajczak}},\ }\href {\doibase 10.1103/PhysRevC.60.034308} {\bibfield
  {journal} {\bibinfo  {journal} {Phys. Rev. C}\ }\textbf {\bibinfo {volume}
  {60}},\ \bibinfo {pages} {034308} (\bibinfo {year} {1999})}\BibitemShut
  {NoStop}%
\bibitem [{\citenamefont {Park}(1995)}]{JMP.1995.36.5453}%
  \BibitemOpen
  \bibfield  {author} {\bibinfo {author} {\bibfnamefont {D.~K.}\ \bibnamefont
  {Park}},\ }\href {\doibase 10.1063/1.531271} {\bibfield  {journal} {\bibinfo
  {journal} {J. Math. Phys.}\ }\textbf {\bibinfo {volume} {36}},\ \bibinfo
  {pages} {5453} (\bibinfo {year} {1995})}\BibitemShut {NoStop}%
\bibitem [{\citenamefont {Jackiw}(1995)}]{Book.1995.Jackiw}%
  \BibitemOpen
  \bibfield  {author} {\bibinfo {author} {\bibfnamefont {R.}~\bibnamefont
  {Jackiw}},\ }\href@noop {} {\emph {\bibinfo {title} {Diverse topics in
  theoretical and mathematical physics}}},\ Advanced Series in Mathematical
  Physics\ (\bibinfo  {publisher} {World Scientific},\ \bibinfo {address}
  {Singapore},\ \bibinfo {year} {1995})\BibitemShut {NoStop}%
\bibitem [{\citenamefont {Goldhaber}(1977)}]{PRD.1977.16.1815}%
  \BibitemOpen
  \bibfield  {author} {\bibinfo {author} {\bibfnamefont {A.~S.}\ \bibnamefont
  {Goldhaber}},\ }\href {\doibase 10.1103/PhysRevD.16.1815} {\bibfield
  {journal} {\bibinfo  {journal} {Phys. Rev. D}\ }\textbf {\bibinfo {volume}
  {16}},\ \bibinfo {pages} {1815} (\bibinfo {year} {1977})}\BibitemShut
  {NoStop}%
\bibitem [{\citenamefont {Kazama}\ \emph {et~al.}(1977)\citenamefont {Kazama},
  \citenamefont {Yang},\ and\ \citenamefont {Goldhaber}}]{PRD.1977.15.2287}%
  \BibitemOpen
  \bibfield  {author} {\bibinfo {author} {\bibfnamefont {Y.}~\bibnamefont
  {Kazama}}, \bibinfo {author} {\bibfnamefont {C.~N.}\ \bibnamefont {Yang}}, \
  and\ \bibinfo {author} {\bibfnamefont {A.~S.}\ \bibnamefont {Goldhaber}},\
  }\href {\doibase 10.1103/PhysRevD.15.2287} {\bibfield  {journal} {\bibinfo
  {journal} {Phys. Rev. D}\ }\textbf {\bibinfo {volume} {15}},\ \bibinfo
  {pages} {2287} (\bibinfo {year} {1977})}\BibitemShut {NoStop}%
\bibitem [{\citenamefont {Grossman}(1983)}]{PRL.1983.50.464}%
  \BibitemOpen
  \bibfield  {author} {\bibinfo {author} {\bibfnamefont {B.}~\bibnamefont
  {Grossman}},\ }\href {\doibase 10.1103/PhysRevLett.50.464} {\bibfield
  {journal} {\bibinfo  {journal} {Phys. Rev. Lett.}\ }\textbf {\bibinfo
  {volume} {50}},\ \bibinfo {pages} {464} (\bibinfo {year} {1983})}\BibitemShut
  {NoStop}%
\bibitem [{\citenamefont {Ganoulis}(1993)}]{PLB.1993.298.63}%
  \BibitemOpen
  \bibfield  {author} {\bibinfo {author} {\bibfnamefont {N.}~\bibnamefont
  {Ganoulis}},\ }\href {\doibase 10.1016/0370-2693(93)91708-U} {\bibfield
  {journal} {\bibinfo  {journal} {Phys. Lett. B}\ }\textbf {\bibinfo {volume}
  {298}},\ \bibinfo {pages} {63} (\bibinfo {year} {1993})}\BibitemShut
  {NoStop}%
\bibitem [{\citenamefont {Davis}\ \emph {et~al.}(1994)\citenamefont {Davis},
  \citenamefont {Martin},\ and\ \citenamefont {Ganoulis}}]{NPB.1994.419.323}%
  \BibitemOpen
  \bibfield  {author} {\bibinfo {author} {\bibfnamefont {A.}~\bibnamefont
  {Davis}}, \bibinfo {author} {\bibfnamefont {A.}~\bibnamefont {Martin}}, \
  and\ \bibinfo {author} {\bibfnamefont {N.}~\bibnamefont {Ganoulis}},\ }\href
  {\doibase 10.1016/0550-3213(94)90045-0} {\bibfield  {journal} {\bibinfo
  {journal} {Nucl. Phys. B}\ }\textbf {\bibinfo {volume} {419}},\ \bibinfo
  {pages} {323} (\bibinfo {year} {1994})}\BibitemShut {NoStop}%
\bibitem [{\citenamefont {Coutinho}\ and\ \citenamefont
  {Perez}(1994)}]{PRD.1994.49.2092}%
  \BibitemOpen
  \bibfield  {author} {\bibinfo {author} {\bibfnamefont {F.~A.~B.}\
  \bibnamefont {Coutinho}}\ and\ \bibinfo {author} {\bibfnamefont {J.~F.}\
  \bibnamefont {Perez}},\ }\href {\doibase 10.1103/PhysRevD.49.2092} {\bibfield
   {journal} {\bibinfo  {journal} {Phys. Rev. D}\ }\textbf {\bibinfo {volume}
  {49}},\ \bibinfo {pages} {2092} (\bibinfo {year} {1994})}\BibitemShut
  {NoStop}%
\bibitem [{\citenamefont {Araujo}\ \emph {et~al.}(2001)\citenamefont {Araujo},
  \citenamefont {Coutinho},\ and\ \citenamefont {Perez}}]{JPA.2001.34.8859}%
  \BibitemOpen
  \bibfield  {author} {\bibinfo {author} {\bibfnamefont {V.~S.}\ \bibnamefont
  {Araujo}}, \bibinfo {author} {\bibfnamefont {F.~A.~B.}\ \bibnamefont
  {Coutinho}}, \ and\ \bibinfo {author} {\bibfnamefont {J.~F.}\ \bibnamefont
  {Perez}},\ }\href {\doibase 10.1088/0305-4470/34/42/310} {\bibfield
  {journal} {\bibinfo  {journal} {J. Phys. A}\ }\textbf {\bibinfo {volume}
  {34}},\ \bibinfo {pages} {8859} (\bibinfo {year} {2001})}\BibitemShut
  {NoStop}%
\bibitem [{\citenamefont {Bonneau}\ \emph {et~al.}(2001)\citenamefont
  {Bonneau}, \citenamefont {Faraut},\ and\ \citenamefont
  {Valent}}]{AJP.2001.69.322}%
  \BibitemOpen
  \bibfield  {author} {\bibinfo {author} {\bibfnamefont {G.}~\bibnamefont
  {Bonneau}}, \bibinfo {author} {\bibfnamefont {J.}~\bibnamefont {Faraut}}, \
  and\ \bibinfo {author} {\bibfnamefont {G.}~\bibnamefont {Valent}},\ }\href
  {\doibase 10.1119/1.1328351} {\bibfield  {journal} {\bibinfo  {journal} {Am.
  J. Phys.}\ }\textbf {\bibinfo {volume} {69}},\ \bibinfo {pages} {322}
  (\bibinfo {year} {2001})}\BibitemShut {NoStop}%
\end{thebibliography}

\end{document}